\documentclass[twocolumn,aps,superscriptaddress,nofootinbib,showpacs,letterpaper]{revtex4}

\usepackage{verbatim}
\usepackage[usenames,dvipsnames]{color}
\usepackage[pdftex]{graphicx}
\usepackage[english]{babel}
\usepackage{amsmath,amssymb,amsfonts}
\usepackage[colorlinks=true,citecolor=blue,linkcolor=magenta]{hyperref}
\usepackage{psfrag}
\usepackage{mathptmx}

\begin{document}

\newcommand{\E}{\mathrm{E}}
\newcommand{\Var}{\mathrm{Var}}
\newcommand{\bra}[1]{\langle #1|}
\newcommand{\ket}[1]{|#1\rangle}
\newcommand{\braket}[2]{\langle #1|#2 \rangle}
\newcommand{\mean}[2]{\langle #1 #2 \rangle}
\newcommand{\be}{\begin{equation}}
\newcommand{\ee}{\end{equation}}
\newcommand{\ba}{\begin{eqnarray}}
\newcommand{\ea}{\end{eqnarray}}
\newcommand{\SD}[1]{{\color{magenta}#1}}
\newcommand{\rem}[1]{{\sout{#1}}}
\newcommand{\alert}[1]{\textbf{\color{red} \uwave{#1}}}
\newcommand{\Y}[1]{\textcolor{yellow}{#1}}
\newcommand{\R}[1]{\textcolor{red}{{\it[#1]}}}
\newcommand{\B}[1]{\textcolor{blue}{#1}}
\newcommand{\C}[1]{\textcolor{cyan}{#1}}
\newcommand{\db}{\color{darkblue}}
\newcommand{\intinfty}{\int_{-\infty}^{\infty}\!}
\newcommand{\Tr}{\mathop{\rm Tr}\nolimits}
\newcommand{\const}{\mathop{\rm const}\nolimits}
\makeatletter
\newcommand{\rmnum}[1]{\romannumeral #1}
\newcommand{\Rmnum}[1]{\expandafter\@slowromancap\romannumeral #1@}
\makeatother

\title{Open quantum dynamics of single-photon optomechanical devices}
\author{Ting Hong}
\affiliation{Theoretical Astrophysics 350-17, California Institute
of Technology, Pasadena, CA 91125, USA}
\author{Huan Yang}
\affiliation{Theoretical Astrophysics 350-17, California Institute
of Technology, Pasadena, CA 91125, USA}
\author{Haixing Miao}
\affiliation{Theoretical Astrophysics 350-17, California Institute
of Technology, Pasadena, CA 91125, USA}
\author{Yanbei Chen}
\affiliation{Theoretical Astrophysics 350-17, California Institute
of Technology, Pasadena, CA 91125, USA}

\begin{abstract}
We study the quantum dynamics of a Michelson interferometer with Fabry-Perot cavity arms and one movable end mirror, and driven by a single photon --- an optomechanical device previously studied by Marshall et al.\ as a device that searches for gravity decoherence. We obtain an exact analytical solution for the system's quantum mechanical equations of motion, including details about the exchange of the single photon between the cavity mode and the external continuum.  The resulting time evolution of the interferometer's fringe  visibility displays interesting new features when the incoming photon's frequency uncertainty is narrower or comparable to the cavity's line width --- only in the limiting case of much broader-band photon does the result return to that of Marshall {\it et al.}, but in this case the photon is not very likely to enter the cavity and interact with the mirror, making the experiment less efficient and more susceptible to imperfections.   In addition, we show that in the strong-coupling regime, by engineering the incoming photon's wave function, it is possible to prepare the movable mirror into an arbitrary quantum state of a multi-dimensional Hilbert space.  
\end{abstract}

\maketitle

\section{Introduction}

Recently, significant progress has been made in observing quantum effects in  macroscopic mechanical systems~\cite{Marquardt}. As presented in the work of O'Connell {\it et al.}\,\cite{connell}, 
a 6-GHz nano-mechanical oscillator was cooled down near its quantum ground state with dilution refrigeration, 
and later prepared into a Fock state by coupling the oscillator to a superconducting qubit. 
States with thermal occupation numbers below unity have also been  achieved with cavity-assisted radiation-pressure cooling, by Teufel {\it et al.}\,\cite{Teufel} and Safavi-Naeini {\it et al.}\,\cite{Painter}.  Further more, as shown by Gupta et al.~\cite{Gupta} and Thompson et al.~\cite{Thompson}, it is possible to couple a single photon strongly with a mechanical degree of freedom, such that the momentum imparted by a single photon to a mechanical degree of freedom can be comparable to its initial momentum uncertainty.  

%These experiments have  demonstrated the experimental feasibility of preparing interesting quantum states, and creating quantum entanglements
%involving macroscopic mechanical degrees of freedom, which will be very important for helping understand the 
%quantum-to-classical transition.

In this paper, we study the open quantum dynamics of a nonlinear optomechanical device, namely a Michelson interferometer with Fabry-Perot cavities, one of them with a movable end mirror (acting as the mechanical oscillator).  This device, driven by a single photon, was proposed by Marshall {\it et al.}~\cite{Marshall,Kleckner} as an experiment to search for Penrose's conjecture of gravity decoherence~\cite{Penrose}.  Such single-photon driven devices have also been more recently studied  by Rabl~\cite{Rabl} and Nunnenkamp {\it et al.}\,\cite{Nunnenkamp}.  By taking advantage of the conserved quantity---the total number of photons in the system, one can obtain exact solutions to this system's quantum dynamics.
Unlike Rabl and Nunnenkamp et al., who studied systematically the statistics of the out-going photons and the steady state reached by the mechanical oscillator, we focus instead on the fringe visibility of a single-photon interferometer, and the conditional quantum state of the mechanical oscillator  upon the detection of an out-going photon.  

\begin{figure}[!h]
\includegraphics[width=3in, bb=0 0 372 220, clip]{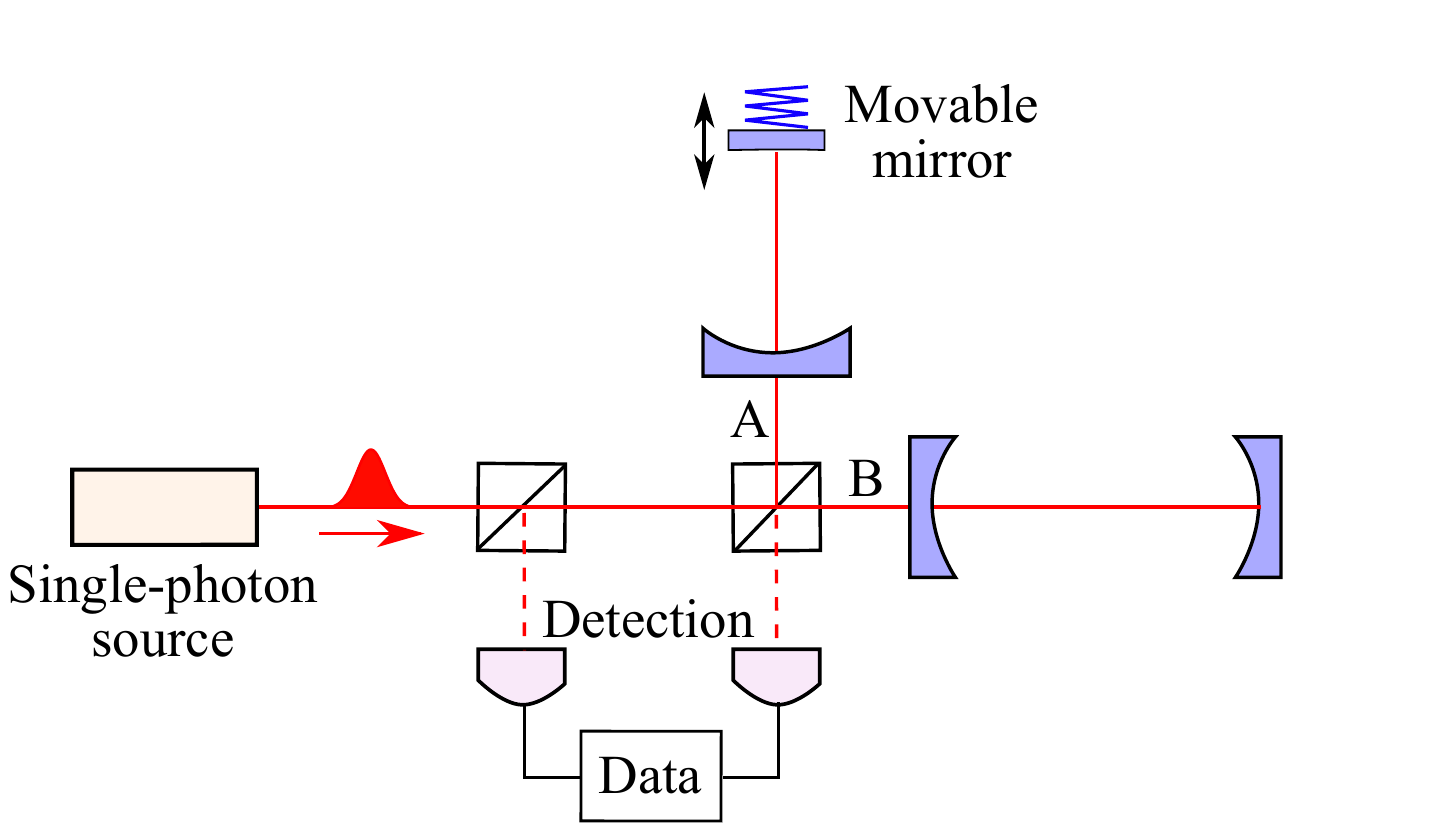}
\caption{(color online) A schematics showing the single-photon interferometer.
The external single photon excites the cavity mode which in turn interacts with the
movable end mirror via radiation pressure. This is adapted from Fig. 1 of Ref.\,\cite{Marshall} 
with small modifications. \label{scheme}}
\end{figure}

The single-photon Michelson interferometer is shown schematically
in Fig.\,\ref{scheme}, in which the port on the left is the input port, towards which the single photon is injected; the photon, after interacting with the Michelson interferometer, may exit either from the input port, or from the other open port. 
Each of the two arms consists of a high-finesse optical cavity; the setup of these two cavities are
identical, except one of them has a movable end mirror, which acts as the mechanical oscillator that
interacts with light in the cavity. The 50/50 beam splitter splits the  quantum state of the entire mirror-light system into two components, one of them corresponding to the photon entering  the 
fixed cavity (and leaveing the oscillator at its initial state), the other corresponding to the photon entering the movable cavity (thereby modifying the 
oscillator's state through radiation pressure). We will set the displacement zero-point of the interferometer to have equal arm lengths, with each arm at a distance equal to the beamsplitter.  At such a zero point, the photon injected from the input port will return to the input port with unit probability.  Therefore we also call the input port the ``bright port'' and the other open port the ``dark port''.  We can artificially tune the interferometer away from its zero point, e.g., by adjusting the fixed microscopic distances between the front mirrors and the beamsplitter.  This changes the relative phase $\varphi$ between the two superimposed components in wave function of the entire system; the resulting variations in the probability density of having the photon exiting the bright port at time $t$, quantified by the fringe visibility, is a measure of the degree of coherence between these two components at this moment in time.

In the case of 
low environmental temperature and in the absence of unexpected mechanisms of decoherence, Marshall
 {\it et al.} showed that the visibility will revive completely for every half of the mechanical
oscillation period. In obtaining such a result, they assumed the photon was initially already in either
of the two cavity arms, and considered a closed evolution of the cavity mode and the mechanical oscillator. 
This assumption has also been widely used in analysis of such a nonlinear optomechanical device, e.g., by Bose
{\it et al.}~\cite{Bose} and subsequent analysis of the Marshall experiment~\cite{Marshall} by Bassi  {\it et al}~\cite{Adler}.

In a realistic experimental setup, it is necessary to take full account of the {\it open quantum dynamics} of this system, which  
that involves the oscillator (the mirror), the cavity mode and the external continuous field, including how the single photon is coupled into the cavity in the first place. 
The open quantum dynamics depends on the wave function of the photon, whose Fourier transform is related to the frequency content of the photon.  For example, if the photon has a short-pulse wave function with time-domain duration much less than the cavity storage time, which corresponds to a  frequency uncertainty  much larger than the cavity line width, then the  photon will only enter the cavity with a small probability.  By contrast, a narrowband photon (with frequency uncertainty below cavity line width) must have a wave packet duration much  
longer than cavity storage time, and therefore we must address the issue that the photon 
can be simultaneously inside and outside the cavity.  The latter scenario, although more complicated, might be 
experimentally more favorable, as in this scenario the photon has a high probability to enter the cavity and to interact with the mirror much more strongly.

The outline of this article goes as follows: in Sec.\,\ref{sec1}, we will write down the Hamiltonian of
our nonlinear optomechanical device and study the open
quantum dynamics by solving the Shr\"{o}dinger equation exactly; in Sec.\,\ref{sec2}, we will 
give a detailed analysis of the single-photon interferometer, and will calculate the interferometer's fringe visibility;
in Sec.\,\ref{sec3}, we will show that the mechanical oscillator can be prepared to an arbitrary 
quantum state in a multi-dimensional Hilbert space, if we inject the single photon with a properly-designed profile into the interferometer; 
in Sec.\,\ref{sec4}, we will summarize our main results.

\section{A Single Cavity with one Movable Mirror}\label{sec1}
\begin{figure}[!h]
\includegraphics[width=1.75in, bb=0 0 178 72, clip]{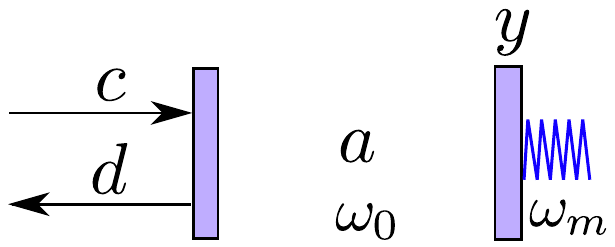}
\caption{(color online) A schematics showing a single-photon interferometer with Fabry-Perot cavity and a movable mirror. 
The displacement of the mirror-endowed mechanical oscillator $y$ is parametrically coupled to the cavity mode
$ a$, which has an eigenfrequency $\omega_0$ with $y=0$. The cavity mode in turn  couples to the ingoing continuous filed $c(x)$ and outgoing continuous field $d(x)$. \label{model}}
\end{figure}

Before studying the entire single-photon interferometer, we first consider  a single cavity, 
as shown schematically in Fig.\,\ref{model}. The cavity has one fixed mirror located at $x=0$, 
and one movable mirror which acts as a mechanical oscillator. Here, assuming the injected photon to have a frequency content much less than the free spectral range of the cavity (which has a relatively high finesse), we will only consider one optical mode of the cavity (which we shall refer to as {\it the cavity mode}). By assuming a high finesse for the cavity, this mode couples to the external vacuum via single-photon exchange.  At linear order in the mirror's motion and assuming low velocity, the coupling between the mirror and the cavity mode is parametric: the position $y$ of the mirror modifies the eigenfreqeuncy of the cavity mode.

\subsection{The Hamiltonian}
\label{hamiltonian}
Here we will write down the Hamiltonian of the system. For simplicity, we will use natural units with $\hbar =1$ and $c=1$ 
throughout this paper. The Hamiltonian of the external continuous optical field, in the position space representation, is given by
\begin{eqnarray}
\hat H_o&=&\frac{i}{2}\int _{-\infty}^0 [(\partial_x \hat c_x^\dag) \hat c_x-\hat c_x^\dag \partial_x\hat c_x]\,{\rm d}x \nonumber\\&+&\frac{i}{2}\int _{-\infty}^0[(\partial_x \hat d_x^\dag) \hat d_x-\hat d_x^\dag \partial_x\hat d_x]\,{\rm d}x
\end{eqnarray}
where $\hat c_x$ and $\hat d_x$ are the annihilation operators for ingoing and outgoing field at location $x$, respectively.
Note that for the actual setup shown in Fig.\,\ref{model}, the ingoing and outgoing field are on the same side of 
the front mirror, namely both at $x<0$. Since the field operators at different locations commute with each other---$[\hat c_{x}\,\hat c^{\dag}_{x'}]=\delta(x-x')$, we can fold the outgoing field from $[-\infty, 0]$ into $[0,+\infty]$, therefore just use $\hat c$ to denote both the
ingoing and outgoing fields, with $\hat c_x(x<0)$ for the ingoing field and $\hat c_x(x>0)$ for the outgoing field, namely
\be
\hat H_o=\frac{i}{2}\int_{-\infty}^{\infty} (\partial_x \hat c_x^\dag \hat c_x-\hat c_x^\dag \partial_x\hat c_x)\,{\rm d}x.
\ee

The free Hamiltonian of the single cavity mode is given by 
 \ba
\hat H_c=\omega_0\hat a^\dag \hat a. 
\ea
with $\hat a$ the annihilation operator and $[\hat a,\,\hat a^{\dag}]=1$. 

The free Hamiltonian for the mechanical oscillator reads
\be
\hat H_m = \frac{ \hat p_y^2}{2m}+\frac{1}{2}m\omega_m^2  \hat y^2 
\ee
where $\hat y$ and $\hat p_y$ are the position and momentum operators, respectively. 

The total interaction Hamiltonian $H_I$ between the external continuum and the cavity mode in the rotating-wave approximation,
and between the cavity mode and the mechanical oscillator, is given by
\be
\label{Hint}
\hat H_I=i\sqrt \gamma(\hat c_{\tiny{0}}\hat a\dag - \hat a \hat c_{\tiny {0}}^\dag)+k\hat a^\dag \hat a \hat y. 
\ee
Here $\gamma=\frac{T}{2L}$ is the cavity bandwidth with $L$ being the cavity length; $k=\omega_0/L$ is 
the optomechanical coupling constant. The interaction between the cavity mode and the external continuum takes place
at the front mirror with $x=0$ and the Hamiltonian describes the exchange of photon between them. The total Hamiltonian 
is a sum of the free and the interaction parts, namely,
\be
\label{Htotal}
\hat H=\hat H_o+\hat H_c+\hat H_m+\hat H_I\,.
\ee
Note once more that by including only a single cavity mode resonant at frequency $\omega_0/(2\pi)$, we must make sure the frequency content of the injected light is focused well within a free spectral range, $c/(2L)$.

\subsection{Structure of the Hilbert Space}

Even though the Hamiltonian contains a cubic term $\hat a^{\dag}\hat a\hat y$, which implies a nonlinear dynamics, we have a conserved dynamical quantity---the total photon number:
\begin{equation}
\hat a^\dagger \hat a + \int_{-\infty}^{+\infty}  \hat c_x^\dagger \hat c_x \, {\rm d}x,
\end{equation}
which makes the system's evolution still analytically solvable, as also recognized by Rabl~\cite{Rabl} and Nunnenkamp et al.~\cite{Nunnenkamp}.  Since the initial state of our system consists of one single photon, there can only be one photon throughout the entire evolution.  Mathematically, this means we only need to consider a one-photon subspace of the entire Hilbert space, which in turn consists of three disjoint subspaces, which corresponds to: $\mathcal{H}_{1-}$, which corresponds to an incoming photon towards the cavity; $\mathcal{H}_2$, which corresponds to a photon inside the cavity, and $\mathcal{H}_{1+}$, which corresponds to a photon leaving the cavity. All quantum states in this space can be written as:
\ba
\label{psi}
\ket{\psi}&=&\int_{-\infty}^{+\infty} f(x,t)e^{-i\omega_0(t-x)}|x\rangle_\gamma\otimes |\phi_1(x,t)\rangle_{\rm m}\,{\rm d}x
\nonumber\\
&+&\alpha(t)e^{-i\omega_0 t} \hat a^\dag|0\rangle_\gamma \otimes |\phi_2(t)\rangle_{\rm m}\,.
\ea
Here
\begin{equation}
|x\rangle_\gamma \equiv \hat c_x^\dagger |0\rangle_\gamma
\end{equation}
is the ``position eigenstate'' of the single photon outside of the cavity, and $|0\rangle_\gamma$ is the optical vacuum; the subscripts $\gamma$ and ${\rm m}$ indicate Hilbert spaces of light and movable mirror, respectively;
$f(x,t)$ is a complex function of position ($-\infty < x<+\infty$) and time, $\alpha(t)$ is a complex function of time $t$ alone;  $|\phi_1(x,t)\rangle_{\rm m}$ and $|\phi_2(t)\rangle_{\rm m}$ are two {\it families of} state vectors that belong to the Hilbert space of the mechanical oscillator.
At any given time, the $x<0$ part of the integral term on the right-hand side corresponds to $\mathcal{H}_{1-}$, the $x>0$ part of the integral term corresponds to $\mathcal{H}_{1+}$, while the non-integral term corresponds to $\mathcal{H}_2$. In general, all three terms will be present, which means the entire system's quantum state is a superposition of having the photon simultaneously present in all three possible locations.   Note that the factors $e^{-i\omega_0(t-x)}$ and $e^{-i\omega_0 t}$ are added to ``factor out'' the free oscillation of the EM field, which has oscillation frequencies near $\omega_0$.  

By imposing normalization conditions of 
\begin{equation}
{}_{\rm m}\langle \phi_1(x,t) |\phi_1(x,t) \rangle_{\rm m}= {}_{\rm m}\langle \phi_2(t) |\phi_2(t) \rangle_{\rm m} =1\,,
\end{equation}
the probability for finding the photon at location $x$ (with $x<0$ indicating a photon propagating towards the cavity, and $x>0$ a photon propagating away from the cavity) is given by
\begin{equation}
p_\gamma(x,t) = |f(x,t)|^2
\end{equation}
while the probability that the photon is in the cavity is given by $|\alpha^2(t)|$. In this way, the normalization condition of the joint quantum state
\begin{equation}
\int_{-\infty}^{+\infty} |f(x,t)|^2 {\rm d}x + |\alpha^2(t)| =1
\end{equation}
is simply a statement about the  conservation of total probability.

The function $f(x,t)$ can be viewed as the out-of-cavity photon's wave function, while $|\phi_1(x,t)\rangle_{\rm m}$ {\it for each $x$} can be viewed as the oscillator state that is entangled with {\it each possibility} for the out-of-cavity photon.  On the other hand, $\alpha(t)$ can be viewed as the probability amplitude of the cavity mode, while $|\phi_2\rangle_{\rm m}$ can be viewed as the oscillator state that is entangled with the in-cavity photon.

To facilitate calculation, for any joint quantum state $|\psi\rangle$, we define
\ba
 \ket{\psi_1(x,t)}_{\rm m}&\equiv& {}_\gamma\langle{x}|{\psi}\rangle e^{i\omega_0(t-x)}=f(x,t) \ket{\phi_1(x,t)}_{\rm m} \quad\\
\ket{\psi_2(t)}_{\rm m}&\equiv &\bra 0 a \ket \psi e^{i\omega_0 t}=\alpha(t)\ket{\phi_2(t)}_{\rm m}
\ea
Here $|\psi_1(x,t)\rangle_{\rm m}$, $-\infty< x <+\infty$, is a series of vectors, parametrized by $x$,  in the Hilbert space of the mechanical oscillator, while $|\psi_2(x,t)\rangle$ is a single vector in the Hilbert space of the mechanical oscillator.  They together carry the full information of the quantum state of the entire system. 
To further appreciate the role of $|\psi_1\rangle_{\rm m}$ and $|\psi_2\rangle_{\rm m}$, we can project each of them into the position eigenstate of the oscillator, $|y\rangle_{\rm m}$, obtaining
\begin{eqnarray}
\Phi_1(t,x,y) \equiv {}_{\rm m}\langle y | \psi_1 \rangle_{\rm m} &=& f(x,t) \phi_1(y,x,t) \\
\Phi_2(t,y) \equiv  {}_{\rm m}\langle y | \psi_2 \rangle_{\rm m} &=& \alpha(t) \phi_2(y,t)
\end{eqnarray}  
which can be viewed as the {\it joint wave functions} of the projection of the entire state into $\mathcal{H}_{1+} \oplus \mathcal{H}_{1-} $ and $\mathcal{H}_2$, respectively. Note that although $f(x,t)$ and $|\phi_1(x,t)\rangle_{\rm m}$ [and similarly $\alpha(t)$ and $|\phi_2(t)\rangle_{\rm m}$] share a phase ambiguity, $|\psi_1(x,t)\rangle_{\rm m}$ and $|\psi_2(t)\rangle_{\rm m}$, and hence $\Phi_1(t,x,y)$ and $\Phi_2(t,y)$ are well defined without ambiguity.

\subsection{Initial, Final States and Photodetection}
\label{subsecdet}

As special cases, we consider the quantum state of the system at $t=0$ (the initial state), and at very late times (the final state).  For the initial state, the photon is propagating towards the cavity, and the cavity is empty.  This corresponds to  $\alpha(0)=0$, and $f(x,0) =0$.  In particular, we also presume the initial state to be {\it separable} between the photon and the oscillator, with
\begin{equation}
|\psi(0) \rangle  = \int e^{i\omega_0 x}F  (x) \ket x_\gamma \,{\rm d}x \otimes|  \phi_0\rangle_{\rm m}
\label{initial}
\end{equation}
Here $F(x)$ is the slowly-varying part of the initial wave function of the photon, and $|\phi_0\rangle_{\rm m} $ the initial wave function of the oscillator.  In other words, we have
\begin{eqnarray}
\label{psiinit}
|\psi_1(t=0)\rangle_{\rm m} &=& F(x) |\phi_0\rangle_{\rm m} \\
|\psi_2(t=0)\rangle_{\rm m}  & = & 0
\end{eqnarray}
with $F(x) =0$ for $x>0$. At a sufficiently  late time $T$, the photon will leave the cavity with unity probability, and we expect $\alpha(T)=0$ and  $f(x,T)=0$ for $x<0$. Mathematically,
\begin{eqnarray}
|\psi_1(x,t \ge T)\rangle_{\rm m} &=& F_{\rm out}(x,t) \ket{\phi(x,t)} _{\rm m}\, \nonumber\\
|\psi_2(t\ge T)\rangle_{\rm m}  & = & 0
\end{eqnarray}
with $F_{\rm out}(t,x)=0$ for $x<0$ and $t>T$. This is an explicitly entangled state between the out-going photon and the mirror, if $|\phi(x,t)\rangle_{\rm m}$ for different values of $x$ are not all proportional to the same state vector.

At an intermediate time $t>0$, suppose a  photodetector is placed at $x=L >0$ (i.e., for out-going photons from the cavity), then the probability density for photon arrival time at $T$ is given by
\begin{equation}
p_L(T) = {}_{\rm m}\langle \psi_1(L,T)| \psi_1(L,T) \rangle_{\rm m}.
\end{equation}
In addition, by detecting a photon at this particular instant, the oscillator is left at a {\it condition quantum state} of $|\phi(x,T)\rangle_{\rm m}$

\subsection{Evolution of the photon-mirror quantum state}

Applying the operations ${}_\gamma\langle x|$ and ${}_\gamma\langle 0| a$ onto the (joint) Schr\"odingier equation
\begin{equation}
\label{schrototal}
i\hbar \frac{{\rm d}|\psi\rangle}{{\rm d}t} =\hat  H |\psi\rangle
\end{equation}
we will obtain coupled equations for $|\psi_1\rangle_{\rm m}$ and $|\psi_2\rangle_{\rm m}$. Throughout this section, we will mostly encounter states in the oscillator's Hilbert space, therefore we will ignore the subscript ``m'' unless otherwise necessary.

\subsubsection{Free Evolution}

For $|\psi_1\rangle$, by applying ${}_\gamma\langle x|$ to both sides of Eq.~\eqref{schrototal} we obtain
\be
\label{psi1}
\left[\partial_t+\partial_x+i\,\hat H_m\right] \ket{\psi_1(x,t)} =-\sqrt{\gamma}\,\delta{(x)}\ket{\psi_2(t)}.
\ee
Equation~\eqref{psi1}, without the $\delta$-function term, simply describes the propagation of the initial photon towards the cavity, and the free evolution of the oscillator. This is because when the single photon is outside the cavity, its propagation is free, while  the oscillator's evolution is unaffected by light.

\begin{figure}
\includegraphics[width=2.5in]{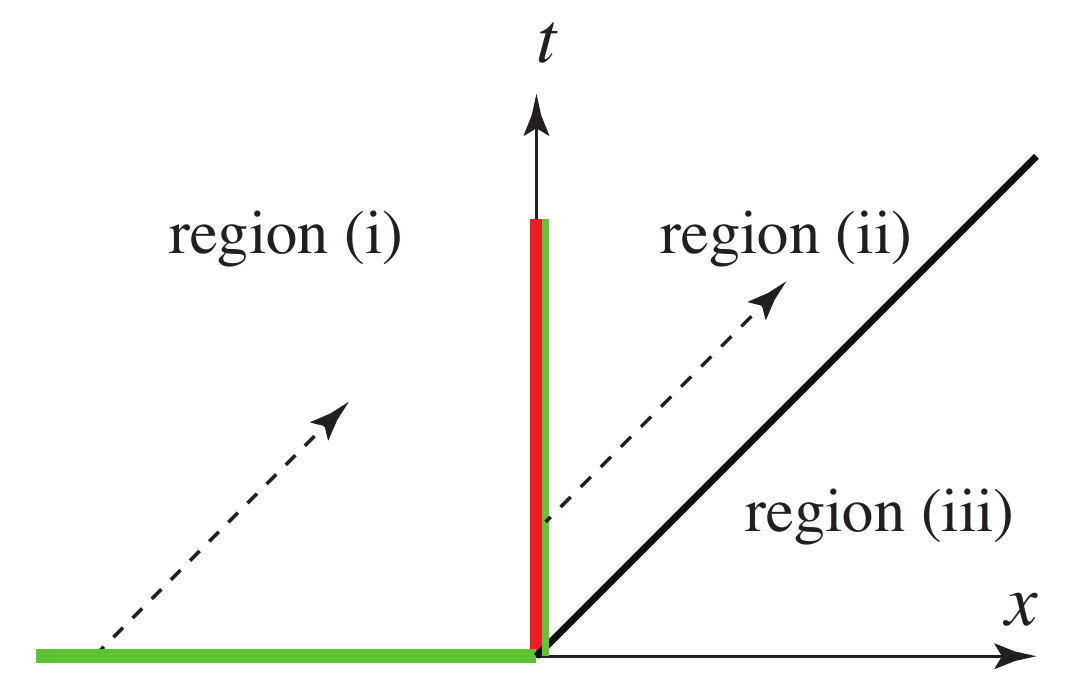}
\caption{(Color Online) Three regions of the $t$-$x$ plane and the free evolutions of $|\psi_1\rangle$. In region (i), the photon has not yet entered the cavity; the joint quantum state of the system is a simple free evolution of the initial quantum state, specified on $t=0$, $x<0$ (green horizontal half line), see Eq.~\eqref{psi1p}.  In region (ii), the photon and the oscillator evolve freely after  propagates after the photon emerges from the cavity; the joint wave function depends on the wave function along $x=0$, $t>0$ (green vertical half line).  The red line diving regions (i) and (ii) corresponds to the $\delta$-function in Eq.~\eqref{psi1}, which embodies the interaction between the outside photon and the in-cavity photon. Region (iii) is causally irrelevant to our experiment.
\label{fig:char} }
\end{figure}

Equation \eqref{psi1}, is a first-order partial differential equation with characteristics along $x-t=\mathrm{const}$. We hereby divide the $t>0$ region of the $t$-$x$ plane into three regions: (i) $x<0$, (ii) $x>0$ and $t>x$, and (iii) $x>0$ and $t<x$, as shown in Fig.~\ref{fig:char}.
We can discard region (iii) right away, because it is not causally connected with our experiment.  In the interiors of regions (i) and (ii) separately, Eq.~\eqref{psi1} has the following general solution,
\begin{equation}
\label{psi1C}
|\psi_1(x,t)\rangle= e^{-\frac{i}{2} \hat H_m (t+x)} |C(t-x)\rangle
\end{equation}
with $|C(v)\rangle$ an arbitrary state-valued function of $v$.

In region (i), $|C(v)\rangle$ can be specified by initial data along the half line of $t=0$, $x<0$; by using 
Eq.~\eqref{psi1C} twice, at $(t,x)$ and $(0,x-t)$,  we obtain [See Fig.~\ref{fig:char}]:
\ba\label{psi1p}
\ket{\psi_1(x<0,t)}= F(x-t) \hat U_m(t) |\phi_0\rangle .\quad
\ea
Here $ U_m$ is the evolution operator for the free oscillator, given by
\begin{equation}
\hat  U_m(t)=e^{-i \hat H_m t}\,.
\end{equation}
In terms of the Fock states $|n\rangle$, we have
\begin{equation}
\hat U_m(t)=\sum_n |n\rangle e^{-i\left(n+\frac{1}{2}\right)\omega_m t} \langle n |
\end{equation}
Equation \eqref{psi1p} corresponds to the photon's wave packet freely propagating along the positive direction of the $x$ axis and the mechanical oscillator independently evolving under its own Hamiltonian.

In region (ii), $|C(v)\rangle$ is specified by boundary data along the half line of $x=0+$, $t>0$, which we denote by
\begin{equation}
|\psi_1(t)\rangle_{0+} \equiv |\psi_1(0+,t)\rangle\,.
\end{equation}
By using Eq.~\eqref{psi1C} twice, at $(t,x)$ and $(t-x,0)$, we obtain \ba
\label{psi1pluspartial}
\ket{\psi_1(x>0,t)}
= \hat U_m (x) |\psi_1(t-x)\rangle_{0+}
\ea
Henceforth in the paper,  $0+$ and $0-$ stand for $x \mapsto 0+$ ($x$ approaches $0$ from positive side of the axis) and $x \mapsto 0-$  ($x$ approaches $0$ from negative side of the axis) respectively. Equation~\eqref{psi1pluspartial} corresponds to the free evolution of the out-going photon and the mechanical oscillator. 

\subsubsection{Junction Condition}

The $\delta$-function on the right-hand side of Eq.~\eqref{psi1} relates the out-going photon to the decay of the in-cavity photon and the reflection of the in-going photon. To take this into account, we simply integrate both sides from $x = 0-$ to $x = 0+$, obtaining:
\be
\label{jump}
\ket {\psi_1(0+,t)}=\ket {\psi_1(0-,t)}-\sqrt \gamma\ket {\psi_2(t)}
\ee
This expresses the out-going wave as a combination of the promptly reflected incoming wave and the wave coming out from the cavity.

\subsubsection{Coupled Evolution}

By applying ${}_\gamma \langle 0 | a$ to both sides of  Eq.~\eqref{schrototal} and using Eq.\,\eqref{jump}, we obtain:
%\begin{align}
%\label{psi2eq}
%&\left[\partial_t+\frac{\gamma}{2} + i\left(\frac{\hat p_y^2}{2m}+\frac{m\omega_m^2 (\hat y-\alpha)^2}{2}-{\beta^2\omega_m}\right)\right]\ket{\psi_2(t)}\nonumber\\
%&\quad~~=\sqrt \gamma \ket{\psi_1(t)}_{0-}
%\end{align}
\ba
\label{psi2eq}
\left[\partial_t +\frac{\gamma}{2} + i\hat H_\gamma\right]|\psi_2(t)\rangle = \sqrt \gamma \ket{\psi_1(t)}_{0-}\,.
\ea
Here as in Eq.~\eqref{jump}, we have defined $\ket{\psi_1}_{0\pm}\equiv \ket{\psi_1( 0\pm,t)}$.  We have also defined 
\begin{equation}
\hat H_\gamma \equiv \frac{ \hat p_y^2}{2m}+\frac{m\omega_m^2 (\hat y-\alpha)^2}{2}-{\beta^2\omega_m}
\end{equation}
with 
\ba
\alpha=-\frac{k}{m\omega_m^2},\quad
\beta=\frac{k}{\omega_m \sqrt {2m\omega_m}}\,.
\ea
The operator $\hat H_\gamma$  can be  viewed as the modified Hamiltonian for the mirror when the photon is present in the cavity.  Here $\alpha$ characterizes the shift in equilibrium position of the harmonic oscillator when the photon is inside the cavity and applies a constant force to the oscillator, while  $\beta$ (as seen from this equation) modifies the eigenfrequency of the harmonic oscillator.  It is easy to work out the eigenstates and eigenvalues of $\hat H_\gamma$: the eigenstates are 
\begin{equation}
\label{shiftfock}
\ket {\tilde n}=e^{i\alpha \hat p_y}\ket n =\hat D(\beta) \ket n
\end{equation} 
which are simply displaced from the original Fock states in phase space, due to the change of equilibrium position, with
\begin{equation}
\hat H_\gamma |\tilde n\rangle = \left(n+\frac{1}{2}-\beta^2\right)\omega_m|\tilde n\rangle 
\end{equation}
which indicates an overall down-shift of eigenfrequency. Here we have further defined the displacement operator
\begin{equation}
\hat D (\beta) \equiv \exp[\beta(b^\dagger -b)] 
\end{equation}
with $b$ and $b^\dagger$ the annihilation and creation operators for the free mechanical oscillator (i.e., before it couples to light). 
  As we shall see in Sec.~\ref{sec4},  $\beta$ will become an important characterizing parameter of our optomechanical device; for example, $\beta\stackrel{>}{_\sim}1$ is the regime in which the device is nonlinear.

For the photon, Eq.~\eqref{psi2eq}  means that the in-cavity photon is continuously driven by the in-coming photon  (right-hand side) and decays towards the out-going photon (as indicated by the $\gamma/2$ term in the bracket on the left-hand side). The above discussion, together with the initial data of $|\psi_2\rangle=0$ at $t=0$ gives
\ba
\label{psi2}
\ket {\psi_2}
=\sqrt{\gamma}\int_0^t e^{-\frac{\gamma}{2}(t-t')}  
\hat U_\gamma(t-t') |\psi_1(t')\rangle_{0-}
\ea
where
\begin{equation}
\hat U_\gamma(t) \equiv e^{-i \hat H_\gamma t} = \sum_n|\tilde n\rangle e^{-i(n+1/2-\beta^2)\omega_m t}\langle \tilde n |,
\end{equation}
which is the modified evolution operator of the oscillator when the photon is in the cavity.

\subsubsection{Full Evolution}

The full evolution of the entire system's quantum state  can now be obtained by combining Eqs.~\eqref{psi1pluspartial}, \eqref{psi1p}, \eqref{jump} and \eqref{psi2}.  In order to study the out-going photon, we only need to consider the region $x>0$ and $t>x$ (see Fig.~\ref{fig:char}), because the emerges from the cavity at $t>0$, and it propagates with $c=1$.   For this region, we obtain a compact-form solution of
\begin{eqnarray}
\label{psi1fin}
|\psi_1(x,t)\rangle &=& \hat{M} |\phi_0\rangle
\end{eqnarray}
where $|\phi_0\rangle$ is the initial quantum state of the oscillator, and
\begin{align}
\label{eq:M}
\hat M &=
\int_0^{t-x} g(t-x,t')   \hat U_m(x)  \hat U_\gamma (t-x-t') \hat U_m(t') {\rm d}t'  \nonumber \\
&=
\int_0^{t-x} g(t-x,t')e^{i\beta^2 \omega_m(t-x-t')} \nonumber \\ 
&\qquad \quad\;\;\;\hat D(\beta e^{-i\omega_m x}) \hat D(-\beta e^{i \omega_m (t'-t)}) \hat U_m(t)  {\rm d}t'\,.
\end{align}
where
\be
g(t,t')\equiv G(t-t')F(-t')
\ee
with
\begin{equation}
G(t) = \delta_+(t) +\gamma e^{-\frac{\gamma}{2} t}
\end{equation}
the cavity's optical Green function. Here the subscript $+$ for the $\delta$ function indicates that it's support lies completely in the region $t>0$. 
Within the operator $\hat M$ [Eq.~\eqref{eq:M}], the factor $g$ contains two terms, the first contains a $\delta$-function and the second an exponential decay over time.  The first term corresponds to the photon being promptly reflected by the cavity's front mirror, while the second term corresponds to the photon staying inside the cavity, for an amount of time equal to $t-x-t'$, which ranges from $0$ to $t-x$.  As a sanity check, it is straightforward to see that when mass of the oscillator approaches infinity, $ \hat U_\gamma$ coincides with $ \hat U_m$, and $\hat M$ simply describes the photon's propagation and the independent evolution of the oscillator.

\section{Single-photon interferometer: Visibility}\label{sec2}

In this section, we will use the results of the previous section to analyze the single-photon interferometer.  

\subsection{The configuration}

We consider a scheme proposed and analyzed by Marshall et al. \cite{Marshall}, which is shown in Fig.~\ref{scheme}.  This Michelson interferometer (with 50/50 beamsplitter) has two arms: in the north arm, the end mirror in cavity A is movable, and initially prepared at a quantum state $|\phi_0
\rangle$,  whereas mirrors in cavity B or east arm are fixed. We assume the photon is injected from the west port, while a fixed photodetector is placed at the south port.  Apart from mirror A being movable, the two cavities are otherwise identical: with the same input-mirror power transmissivity $T$, length $L$ (for cavity A, counted from the zero-point of A's displacement).  The front mirrors are placed at equal macroscopic distance from the beamsplitter, while there is a phase detuning of $\varphi$ in arm B for $\omega_0$~\footnote{To give rise to a detuning, we assume that all optical frequencies we consider are centered around $\omega_0$, and we offset the location of cavity $B$ from symmetry by a length $l$ such that $\omega_0 l=\varphi/2$. }.   In our convention, if mirror A is at zero point and $\varphi=0$, the photon will always return to the west port. Henceforth in the paper, we shall refer to the west port as the input port, and the south port the output port --- although we may not always find the photon at the output port.  Indeed, whether and when the photon arrives at the photodetector is jointly determined by $\varphi$ and the state of motion of mirror A.

In particular, we shall use
$p(t)$ to denote the probability density for the photon to arrive at the detector at $t$ (which can be measured by repeating the experiment many times).  If we idealize the arrival time of the in-going photon (at the front mirror) to be $t=0$, and ignore the macroscopic distance between the front mirrors, the beamsplitter, and the photodetector, then we are interested in $p(t)$ at $t\ge 0$. We further define an {\it instantaneous fringe visibility}
\begin{equation}
\label{vinst}
v(t) = \frac{p_{\max}(t)-p_{\min}(t)}{p_{\max}(t)+p_{\min}(t)}\,,
\end{equation}
which measures the degree of coherence between the two components of returning photons at the beamsplitter, and can only become unity if at time $t$ the joint mirror-photon quantum state is separable, as we shall see more clearly in Sec.~\ref{subsec:final}.

\subsection{The role of the beamsplitter and a decomposition of field degrees of freedom}

In Sec.~\ref{subsecdet}, we have studied in detail how the photon first affects the $x<0$ components of the optical field out-side of a cavity, then interacts with the mirror, and finally returns back to the $x>0$ components of the optical field.  The scenario for a Michelson interferometer is slightly more complicated: we now need to consider a set of {\it input fields} that replaces the $x<0$ single field in the single-cavity case, and a set of {\it output fields} which replaces the $x>0$ single field.

\begin{figure}
\includegraphics[width=3.25in, bb=0 0 372 220, clip]{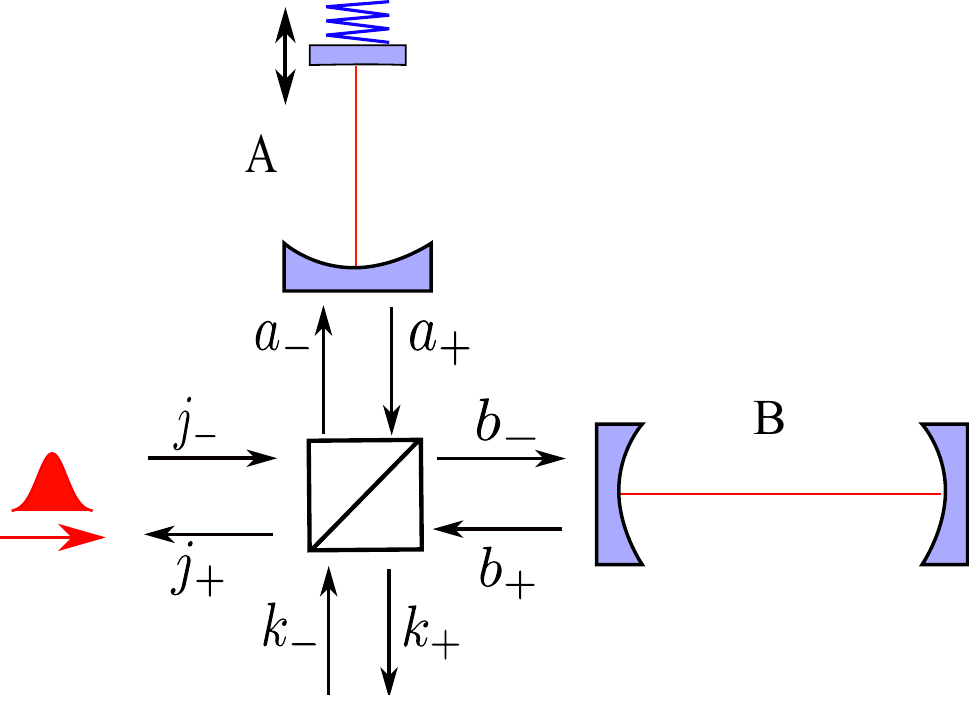}
\caption{We illustrate the fields entering and exiting each of the four ports of the interferometer.  We use arrows to define the positive sense of the coordinate used to label their locations.  For each of them $x=0$ corresponds to the location of the beamsplitter.
\label{scheme:bs}}
\end{figure}

As shown in Fig.~\ref{scheme:bs}, the annihilation operators of the input field for the two cavities are ($\hat j_-$, $\hat k_-$, $\hat a_-$, $\hat b_-$), while those of the output fields for the cavities are ($\hat j_+$, $\hat k_+$, $\hat a_+$, $\hat b_+$).  Each of these files are defined as a function of $-\infty < x<  +\infty$, with $x=0$ corresponding to the position of the beamsplitter, and positive direction along the arrow shown in Fig.~\ref{scheme:bs}.  Ultimately, we need to calculate the fields of $\hat j_+$ and $\hat k_+$ in terms of $\hat j_-$ and $\hat k_-$.

Note that at by allowing $x$ to run through the entire real axis, we have assigned {\it two} input fields and two output fields to each point along the optical path (note here that ``input'' and ``output'' refer to the cavities, {\it not the beamsplitter}).  This {\it redundancy} is necessary for a simplified treatment of the beamsplitter: instead of treating its internal dynamics, we simply view it as a mapping between the two different {\it representations} of the input and output fields.  One representation ($\hat j_{\pm}$, $\hat k_{\pm}$) corresponds to the point of view of observers at the west and south ports, pretending that the beamsplitter does not exist; the other ($\hat a_{\pm}$, $\hat b_{\pm}$) corresponds to the point of view of observers at the east and north ports. 

The conversion between the two representations takes the same form as the ``input-output relation'' of the beamsplitter:
\begin{equation}
\label{map}
\hat b_\pm(x)=\frac{\hat j_\pm(x)-\hat k_\pm(x)}{\sqrt 2}\,,\; \hat a_\pm(x)=\frac{\hat j_\pm(x)+\hat k_\pm(x)}{\sqrt 2}\,.
\end{equation}
As an example, consider a quantum state in which a (instantaneous) photon is injected from the input port, which, according to the mapping in Eq.~\eqref{map}, has two equivalent representations:
\begin{equation}
\label{photonexample}
 \hat j_-^\dagger(x_0) |0\rangle = \frac{\hat a_-^\dagger(x_0)+\hat b_-^\dagger(x_0)}{\sqrt{2}} |0\rangle\,.
\end{equation}
As time grows, the quantum state evolves as $x_0 \rightarrow x_0+t$. At any instant, the left-hand side represents a single photon propagating from west to east, and continue through the location of the beamsplitter.  The right-hand side represents a photon has a two-component wave function, the first component propagates northwards, the second eastwards.

Although the two representations are equivalent, we still prefer to use the south-west representation when treating the generation and detection of  photons, and the north-east representation when treating the light's interaction with the cavities.

\subsection{Interactions between light and cavities}
For each individual cavity, we intend to apply the result of Sec.~\ref{hamiltonian}.  We note that $\hat a_-(x)$ (for $x<0$) and $a_+(x)$ (for $x>0$) defined in this section maps to the $\hat c(x)$ (for $x<0$) and $\hat d(x)$ (for $x>0$) respectively, as defined in Sec.~\ref{hamiltonian} and illustrated in Fig.~\ref{model}. For this reason, we define
\begin{equation}
\hat a(x) \equiv \left\{
\begin{array}{cl}
\hat a_-(x)\,, & x<0\,, \\
\\
\hat a_+(x)\,, & x>0\,,
\end{array}\right.
\label{ax}
\end{equation}
and
\begin{equation}
\hat b(x) \equiv \left\{
\begin{array}{cl}
\hat b_-(x)\,, & x<0\,, \\
\\
\hat b_+(x)\,, & x>0\,.
\end{array}\right.
\label{bx}
\end{equation}
In this way, $a(x)$ and $b(x)$ here both map to $c(x)$ defined in Sec.~\ref{hamiltonian}.  [The $a$ and $b$ here are not to be confused with operators of the optical mode and the mechanical oscillator ---  we shall always explicitly include the argument $(x)$ for these continuum operators.]  We further define
\begin{equation}
\hat j(x) \equiv \left\{
\begin{array}{cl}
\hat j_-(x)\,, & x<0\,, \\
\\
\hat j_+(x)\,, & x>0\,,
\end{array}\right.
\label{jx}
\end{equation}
and
\begin{equation}
\hat k(x) \equiv \left\{
\begin{array}{cl}
\hat k_-(x)\,, & x<0\,, \\
\\
\hat k_+(x)\,, & x>0\,.
\end{array}\right.
\label{kx}
\end{equation}
Furthermore, for fields $a$, $b$, $j$ and $k$, the transformation relations Ea.~\eqref{map} also apply.

Now suppose at $t=0$, we have a photon coming from the input (west) port with arbitrary wave function $F(x)$ [like in Eq.~\eqref{initial}, here $F(x)=0$ for $x>0$]. The initial quantum state of the entire optomechanical system is
\ba
\label{eqn:t0}
\ket {\psi(0)}&=&\int_{-\infty}^0 {\rm d}x\,F(x)\hat j^\dag_-(x) \ket 0_\gamma\otimes {\ket{ \phi_0}}_A
\ea
Since we would like to investigate this state's evolution when the photon reaches the cavity, we covert into the north-east representation:
\ba
\ket {\psi(0)}=\frac{1}{\sqrt{2}}\left[ |{\psi_A(0)}\rangle + |{\psi_B(0)}\rangle\right]
\label{psidecomp}
\ea
Here we have defined
\ba
|{\psi_A(0)}\rangle &=&   \int_{-\infty}^0 {\rm d}x\,F(x)\hat a^\dag(x) \ket 0_\gamma\otimes {\ket{ \phi_0}}_A \\
|{\psi_B(0)}\rangle &=& \int_{-\infty}^0 {\rm d}x\,F(x)\hat b^\dag(x) \ket 0_\gamma\otimes {\ket{ \phi_0}}_A
\ea
in which we have already taken Eqs.~\eqref{ax} and \eqref{bx} into account.

Here $|\psi_A(0)\rangle$ corresponds to the case in which the photon enters Cavity A with the movable mirror, and $|\psi_B(0)\rangle$ the case in which the photon enters Cavity B with the fixed mirror. As time goes on, these two states evolve individually, and Eq.~\eqref{psidecomp} remains true for $t>0$.  For the Cavity A component $|\psi\rangle_A$, we have [cf. Eq.\,\eqref{psi1fin}]
\begin{align}
|\psi_A(t)\rangle = \int_0^{t-x}{\rm d}t'& g(t-x,t') e^{i\beta^2\omega_m(t-x-t')} \hat D(\beta e^{-i\omega_m x})  \nonumber\\
& \hat  D (-\beta e^{i\omega_m(t'-t)})\hat  a^\dagger(x) |\Phi(t)\rangle
\end{align}
where we have defined
\begin{equation}
|\Phi(t) \rangle \equiv  \hat U_m(t)|0\rangle_\gamma |\phi_0\rangle_A\,,
\end{equation}
while for $|\psi\rangle_B$, we set $\beta\rightarrow 0$ and obtain
\begin{equation}
|\psi_B(t)\rangle = e^{i\varphi}\int_0^{t-x}{\rm d}t'g(t-x,t')\hat   b^\dagger(x)  |\Phi(t)\rangle.
\end{equation}

\subsection{The final state}
\label{subsec:final}
In order to describe the quantum state seen by the photodetector,  we map $a$ and $b$ into $j$ and $k$, only keeping the $k$ component.  We further project onto the single-photon basis of ${}_\gamma\langle 0|k(x)$, assuming $x=0+$, obtaining
\begin{equation}\label{eqcomb}
|\psi(t)\rangle_\mathrm{m}=\frac{1}{2}\left[|\psi_A(t)\rangle_{\rm m} +e^{i\varphi} |\psi_B(t)\rangle_{\rm m}\right]
\end{equation}
with
\ba
|\psi_A(t)\rangle_\mathrm{m} &=&\int_0^t {\rm d}t' g(t,t')\hat  O(t-t')|\phi_0(t)\rangle \\
|\psi_B(t)\rangle_\mathrm{m} &=&\int_0^t {\rm d}t' g(t,t')|\phi_0(t)\rangle \label{eqarmb}
\ea
with
\begin{equation}
|\phi_0(t)\rangle \equiv  \hat U_m(t) |\phi_0\rangle
\end{equation}
and
\be
 \hat O(t) \equiv e^{i\beta^2\omega_m t} \hat  D(\beta)  \hat D(-\beta e^{-i\omega_m t})\,,
\ee
in particular $ \hat O(0) = 1$.  In this way, we are using the same notation as Eq.~\eqref{psi1fin}, and we can use Sec.~\ref{subsecdet} for obtaining photo-detection probability density at each time $t>0$, which is given by
\begin{equation}
p(t) =\frac{\||\psi_A\rangle_{\rm m} \|^2+\||\psi_B\rangle_{\rm m} \|^2+2\mathrm{Re}\left(e^{i\varphi}{}_{\rm m}\langle\psi_A|\psi_B\rangle_{\rm m}\right) }{4}
\label{eqn:probability}
\end{equation}
which, when adjusting values of $\varphi$,  leads to an instantaneous visibility of [Cf.~\eqref{vinst}]:
\begin{equation}
v(t) =\frac{2|_{\rm m}\langle\psi_A|\psi_B\rangle_{\rm m}|}{\||\psi_A\rangle_{\rm m} \|^2+\||\psi_B\rangle_{\rm m} \|^2 }.
\label{eqn:visibility}
\end{equation}
It relies on how different $\psi_A$ is from $\psi_B$, which indicates how much the movable mirror in Cavity A is capable of ``learning'' about the existence of the photon in Cavity A. At any instant, if $\psi_A$ is proportional to $\psi_B$ (differ by a phase), the state of the movable mirror does not change, and therefore we have a perfect visibility.  By contrast, if the photon is able to transform the movable mirror into a state substantially different from its freely evolving state, e.g., the orthogonal state in the extreme case, then we will have a significantly
reduced visibility.

Similar to Eq.~\eqref{psi1fin}, here $\psi_A$ and $\psi_B$ each has a promptly reflected part [which arises from the $\delta$-function part of $g(t,t')$], and a part in which the photon enters the cavity [which arises from the exponential decay part of $g(t,t')$].  It is the second part that contributes to the reduction of visibility.

\subsection{Examples}

We consider an experimental situation with the central frequency of the injecting photon tuned to the resonant frequency of the cavity, with a wave function of 
\be
F(x) = \sqrt{2 \Gamma} e^{\Gamma x}\Theta(-x)\,.
\ee
Here $\Gamma$ measures the frequency-domain width of the photon.  We further assume that the mechanical oscillator's eigenfrequency (when uncoupled with light) is equal to the cavity bandwidth, or $\omega_m=\gamma$.  As in Ref.\,\cite{Marshall}, we assume that the mechanical oscillator, i.e., the mirror, is initially prepared at 
its ground state:
\be
\ket{\phi_0}=\ket {0}_A.
\ee
With these specializations, we have
\ba
\label{psiAhere}
&\ket{\psi_A(t)}_{\rm m}& = C(t)\left[\ket 0+\gamma  \ket {M(t)}\right] \\
\label{psiBhere}
&\ket{\psi_A(t)}_{\rm m}& =C(t)\left[\ket 0+\gamma \int^t_0 dt'\, f(t-t')\ket 0 \right]
\ea
with 
\begin{equation}
C(t)  \equiv \sqrt{2 \Gamma} e^{-(\Gamma+i\omega_m/2)t} \,,\quad
f(t) \equiv e^{(\Gamma-\gamma/2)t } \end{equation}
%By using the above result Eq.~(\ref{eqcomb}),  we obtain the following {\it unnormalized} conditional quantum state of the mechanical
%oscillator with a photon detected at dark port at time $t$:
%\ba
%\label{psimhere}
%\ket {\psi}_{\rm m} ={}_{\gamma} \bra {0} \hat k_+ \ket \psi =\frac{1}{2}(\ket{\psi}_A+e^{i\varphi} \ket{\psi}_B)\,,
%\ea
and 
\ba
%\ket M\equiv \int^t_0 dt' \, f(-t')e^{i\beta^2\omega_m (t-t')}D(\beta)D(-\beta e^{i\omega_m(t'-t)})\ket 0 \\
\ket {M(t)} \equiv \int^t_0 dt' \, f(t-t')e^{i\beta^2\left[\omega_m(t-t')-\sin\omega_m(t-t')\right]} \nonumber \\
\times \left |\beta-\beta e^{i\omega_m(t'-t)}\right\rangle
\ea
By comparing with Sec.~\ref{subsec:final}, we first find that visibility depends on the similarity between $|M(t)\rangle$ and its counterpart in Eq.~\eqref{psiBhere}: when they are similar to each other (e.g., when $\beta \stackrel{<}{_\sim} 1$ ) or when they do not contribute significantly to $|\psi_{A,B}(t)\rangle_m$, the visibility will tend to be high.  By contrast, in order to achieve a complete incoherence, we need $|M(t)\rangle$ to contribute significantly, and nearly orthogonal to $|0\rangle$ --- and this {\it requires} $\beta \stackrel{>}{_\sim} 1$.  The arrival probability density \eqref{eqn:probability}  and  contrast defect \eqref{eqn:visibility} can be computed if we use
\begin{equation}
\langle 0 |\beta\rangle = \langle 0 |\hat D(\beta) |0\rangle =e^{-\beta^2/2}
\end{equation}

%
%an alternative
%\begin{equation}
%|\psi\rangle_A = e^{-(a+i\omega_m/2 ) t }\left[ |0\rangle + 
%|M_A\rangle \right]
%\end{equation}
%with
%\begin{equation}
%|M_A\rangle = \gamma \int_0^t  dt' e^{(a-\gamma/2)(t-t')} 
%\end{equation}

%Inserting our particular expressions \eqref{psimhere}--\eqref{psiBhere} into Eqs.~\eqref{eqn:probability} and \eqref{eqn:visibility},  for various choices of $\gamma$ and $\beta$, we obtain the maximum/minimum detection probability density (when varying $\varphi$), as well as visibility, as function of $t$.  Here we have fixed $\gamma=\omega_m=1$.  

\begin{figure}
\includegraphics[width=0.48\textwidth]{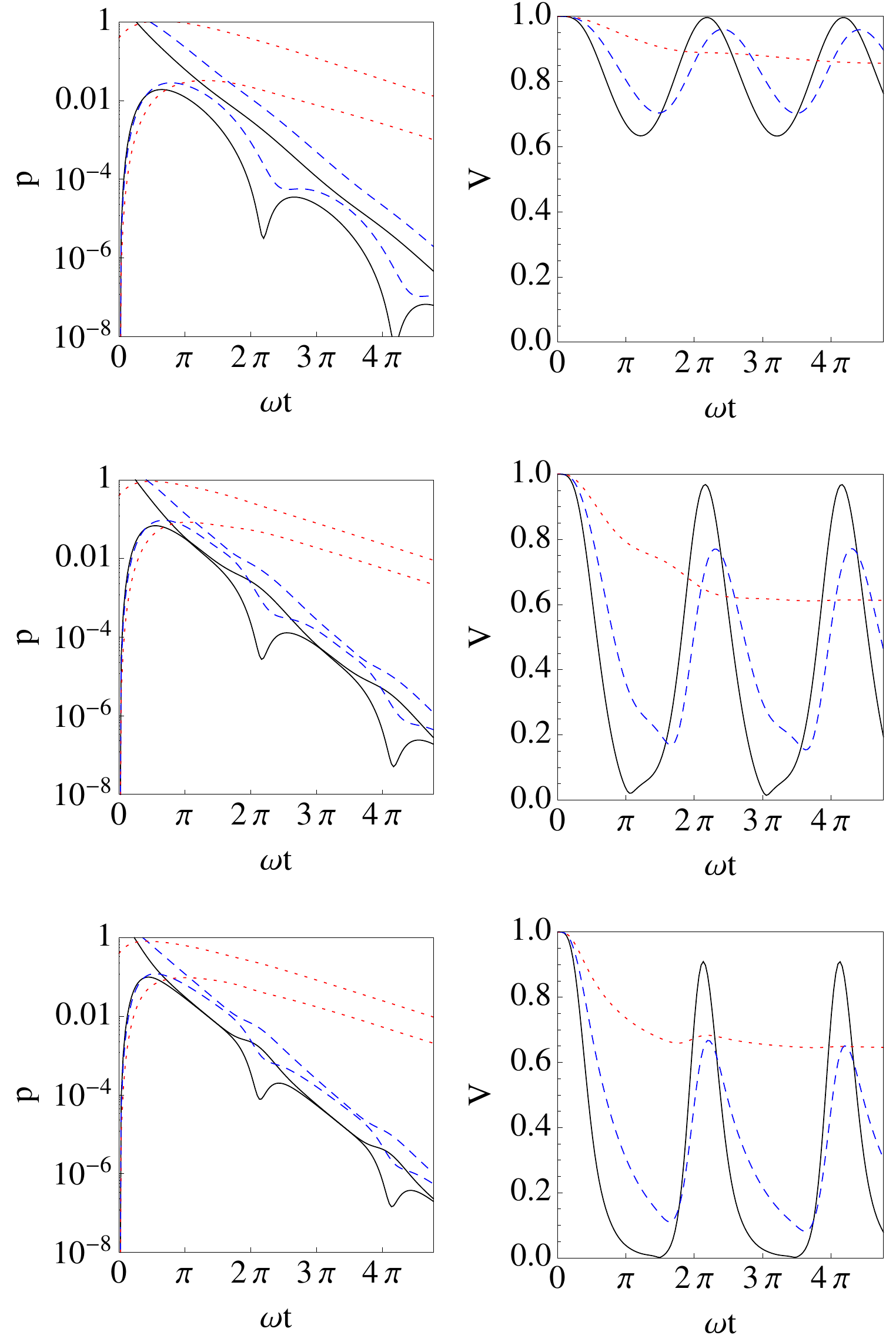}
\caption{(color online) (left) Probability density and (right) fringe visibility for the photon to come out with different $\beta$: (Top-to-bottom: first row, $\beta=0.5$;
second row, $\beta=1.2$; third row, $\beta=2$). For each $\beta$, three different values of $\Gamma$ are considered for comparison: $\Gamma=0.2$ (red dotted), 1(blue dashed), 2(black solid). All the calculation assume $\gamma=1, \omega_m =1$. For probability density plot, the upper line of the same color is the maximum value of the probability density, the lower one is the minimum value.\label{fig:visi_p}}
\end{figure}

In Fig.~\ref{fig:visi_p}, we plot maximum and minimum of the probability density in the left panels, and visibility in the right panels, both as functions of time.  We have chosen $\beta=0.5$ for upper panels, $\beta=1.2$ for middle panels and $\beta=2$ for lower panels.  In each panel, we have also shown curves with $\Gamma=0.2$ (red dotted), $\Gamma=1$ (blue dashed) and $\Gamma=2$ (solid black).   As $\beta$ increases (as we move from upper to lower panels), the photon's ponderomotive effect on the movable mirror increases, therefore the visibility is able to vary more. This means $\beta\stackrel{>}{_\sim} 1$ is necessary (but not sufficient, see below) for visibility to substantially decay and then revive --- a feature Ref.~\cite{Marshall} has used to search for decoherence effects. 

On the other hand, another condition for visibility to first decrease and then revive, and repeat on, seems to be $\Gamma\stackrel{>}{_\sim} 1$, as also indicated by each of the right panels of Fig.~\ref{fig:visi_p}.  In addition, as $\Gamma \gg 1$, our result becomes comparable to Ref.~\cite{Marshall}.  Qualitatively, this is because for $\Gamma\gg 1$, if photon does arrive at a time around $\sim 1$, we can be sure the photon has interacted with the mirror --- and we can roughly treat the photon as already within the cavity at $t=0$.  
%The probability density for detection of a photon at the dark port at time $t$ is
%\ba
%p(t)&=&  |{}_\gamma\langle 0 |\hat k_+ |\psi(t)\rangle|^2 ={}_{\rm m} \langle \psi| \psi\rangle_{\rm m}. 
%\ea

Mathematically,  for $t \gg 1/\Gamma$,  the conditional quantum state of the mirror given photon detection at time $t$ could be approximately written as : 
\ba
\ket{\psi}_{\rm m} &=& \frac{\gamma}{\sqrt{2a}}e^{-(\gamma+i\omega_m)t/2}\nonumber\\
&&\left[e^{i\varphi} \ket 0
+e^{i\beta^2\omega_m t}\hat D(\beta)\left |-\beta e^{-i\omega_m t}\right\rangle\right].\quad \\
&=&\frac{\gamma}{\sqrt{2a}}e^{-(\gamma+i\omega_m)t/2}\nonumber\\
&&\left[e^{i\varphi}\ket 0
+ e^{i\beta^2(\omega_m t-\sin{\omega_m t}) }\left |\beta-\beta e^{-i\omega_m t}\right\rangle\right].\quad\quad
\ea
This is consistent with results of Ref.~\cite{Marshall}.  

However, in order for $a \gg  1$ and to observe a revival of visibility, we have to wait till $t \ge  2\pi$. The probability for detecting the photon at such late times is exponentially small --- as indicated by the left panels of Fig.~\ref{fig:visi_p}  This means we may have to make a trade off between having a very sharp revival of visibility and being robust against loss and able to cumulate enough statistics within reasonable amount of time.

\section{Conditional Quantum-state preparation}\label{sec3}

In this section, we show how to engineer an arbitrary quantum state of the mechanical oscillator by injecting a single photon with specifically designed wave function and by post selecting the arrival time of the output photon. Note that unlike Refs.~\cite{Rabl,Nunnenkamp}, our state preparation procedure is conditional.  This guarantees a pure quantum state for the mechanical oscillator, but requires a low decoherence rate and a high detection quantum efficiency for the out-going photon. 

\subsection{The configuration}

\begin{figure}
\includegraphics[width=0.4\textwidth, clip]{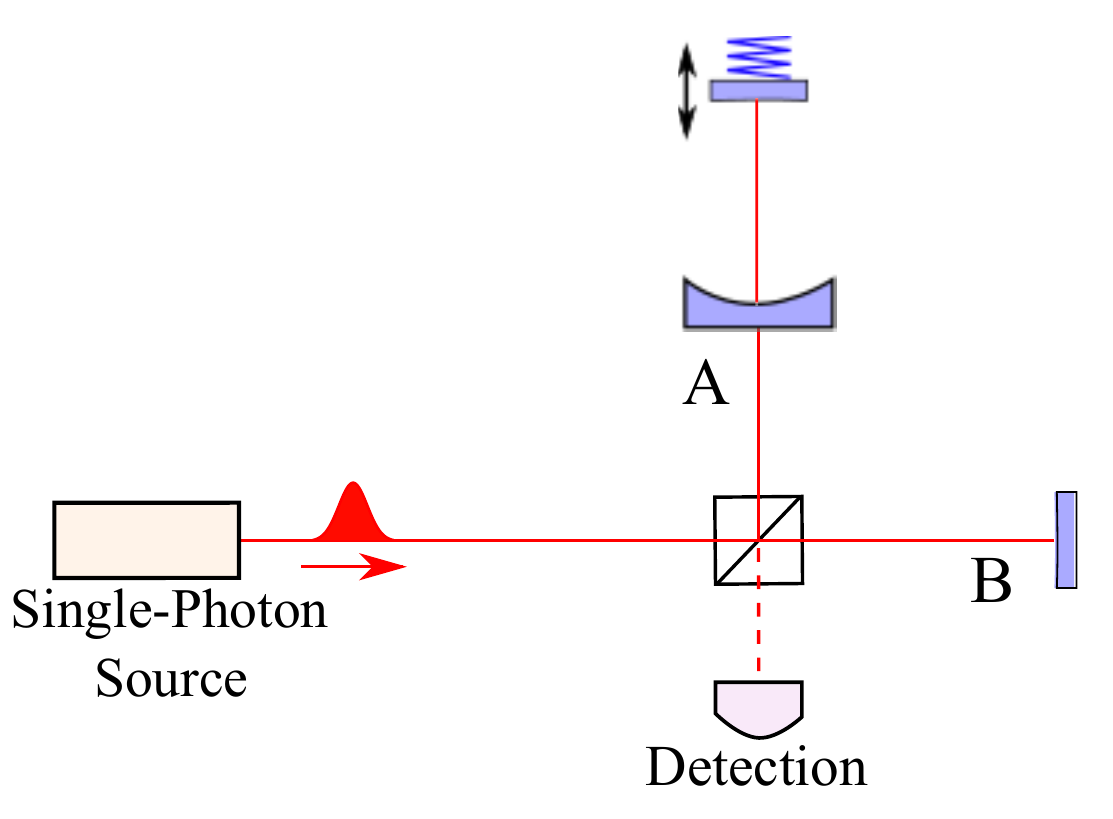}
\caption{ (color online) The sample device which uses single photon to prepare mechanical oscillator quantum state. Here the the detuning phase for the mirror on the east arm is adjusted such that the promptly reflected photon will come out from west port, with $0$ probability coming out from south port.\label{scheme2}}
\end{figure}

The scheme is shown in Fig.~\ref{scheme2}.  It is very similar to the single-photon interferometer discussed in the previous section, except that in the east arm we replace the cavity B with a perfectly reflected mirror B. In this case, most of the previous analysis are still valid: Eq.~(\ref{eqcomb}) to Eq.~(\ref{eqarmb}). The only difference is that the $g(t,t')$ function in Eq.~(\ref{eqarmb})
%\R{Please always put a space between ``Eq.''\ and (...)!!}
 needs to be replaced by $\delta(t-t')$, as we have a perfectly reflecting mirror instead of a cavity here, namely,
\begin{equation}
\ket{\psi_B(t)}_{\rm m} = \ket{\phi_0(t)}
\end{equation}

To proceed, we further adjust the detuning phase $\varphi$ in Eq.~(\ref{eqcomb}) such that at the dark port, the promptly reflected  wave from the front mirror of cavity A exactly cancels the promptly reflected wave from the mirror B. In this case, having a photon emerging from our detection port (Fig.~\ref{scheme2}) automatically indicates that the photon has entered the cavity and interacted with the mirror;  Eq.~(\ref{eqcomb}) or the conditional quantum state of the mechanical oscillator (unnormalized) is given by:
\be\label{eqsa}
\ket{\psi(t)}_{\rm m} = \frac{1}{2}\int_0^t {\rm d}t'\, g_p(t,t')\hat {O}(t-t')\ket{\phi_0(t')}
\ee
with
\be
g_p(t,t')=\gamma e^{-\gamma/2(t-t')}F(-t'). 
\ee
 As $g_p(t,t')$ is related to the input photon wave function $F(x)$, by modifying input photon wave function, we can therefore engineer the conditioning mechanical oscillator quantum state $\ket{\psi(t)}_m$. Even if there is a finite probability that the 
 photon will come out through the west arm or the bright port, once we detect a photon at time $t$ at the dark port, we know that it must come from arm A and it also has stayed in the cavity A for a certain amount of time.

\subsection{Preparation of a single displaced-Fock state}

First of all, we notice that when different in-coming photon wave function $F$'s are used, if we keep conditioning over the {\it same} photon arrival time $t$, the conditional quantum states we obtain for the mechanical oscillator will depend linearly on $F$.  In other words,  if $F_1$ allows us to prepare $|\phi_1\rangle$,  and $F_2$ allows us to prepare $|\phi_2\rangle$, then injecting a new photon with a superimposed wavefunction $F= \alpha_1 F_1+ \alpha_2 F_2$ will allow us to prepare $\alpha_1 |\phi_1\rangle+ \alpha_2|\phi_2\rangle$

This means we only need to show how members of a complete basis can be prepared, and we choose this to be 
\be\label{eqdpn}
\ket{\psi(t)}_{\rm m} =|\tilde n\rangle =  \hat D(\beta)\ket{n}\,,\quad n =0,1,2\ldots\,.
\ee
These displaced Fock states are simply Fock states of the oscillator when the photon is inside the cavity, see Eq.~\eqref{shiftfock}.

Let us assume that the mechanical oscillator is initially prepared at its ground state. 
Before studying preparation of an arbitrary conditional quantum state for the mechanical oscillator,  
we first show that we can prepare a conditional state with an arbitrary quantum number $n$,
by injecting a photon with the following wave function:
\be\label{eqdn}
F(x) =\sqrt{\gamma} e^{(\gamma/2-i\beta^2 \omega_m + i n \omega_m)x}\Theta(-x) .
\ee
%Here $D(\beta)$ is the displacement operator defined in Eq.(\ref{eqdispl}).

\begin{figure}
\includegraphics[width=0.4\textwidth]{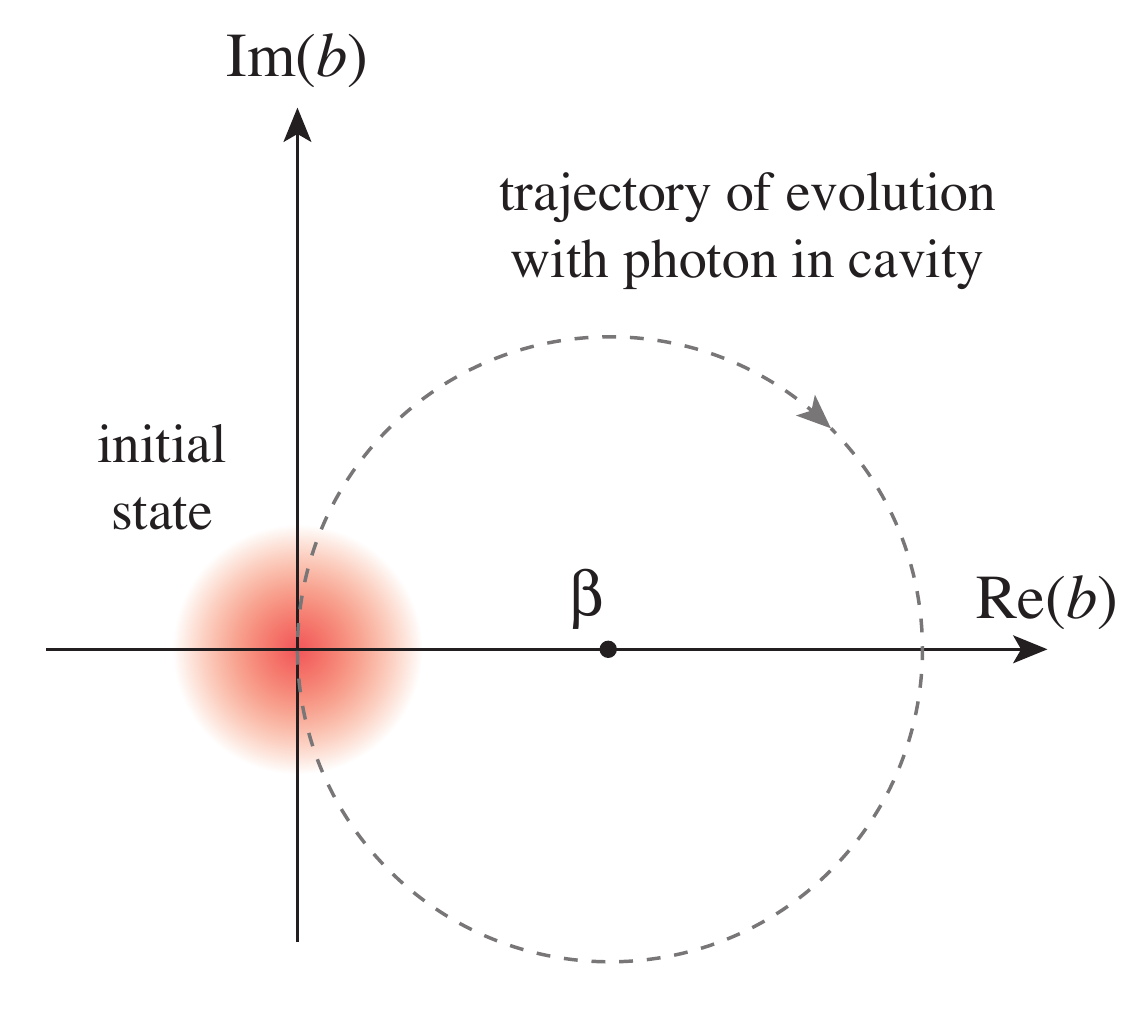}
\caption{A sketch of the phase-space trajectory of the mechanical oscillator. The Wigner function of the initial state $|0\rangle$ is represented by the shaded disk, the dot marked with $\beta$ on the real axis is the new equilibrium position of the oscillator when the photon is in the cavity, while the dashed circle is the trajectory of the oscillator's Wigner function when the photon is inside the cavity. Detection of the out-going photon at $t=2n\pi/\omega_m$ corresponds to superimposing all mechanical-oscillator quantum states along the dashed trajectory, weighted by the photon's wave function. \label{fig:phasespace}}
\end{figure}

As we plug Eq.(\ref{eqdn}) into Eq.(\ref{eqsa}) we obtain the 
conditional quantum state of
\be\label{eqcft}
|\psi(t)\rangle_m =\frac{\hat D(\beta)\gamma^{3/2}e^{-\frac{\gamma}{2}t +i\beta^2 \tau}}{2\omega_m}\int_0^\tau d\tau'
e^{-i n \tau'}\ket{-\beta e^{i (\tau'-\tau)}}
\ee
with $\tau \equiv \omega t$.   This is a coherent superposition coherent states, which in the complex amplitude domain all line up in a circle with radius $\beta$ around the center located at complex amplitude equal to $\beta$; these states are parametrized mathematically by $\hat D(\beta) |-\beta e^{i\phi}\rangle$.
These states are superposed with the same magnitude, but different phases, due to the decay rate of $\gamma/2$ in the $F$ chosen by Eq.~\eqref{eqdn}. Obtaining such a state is understandable, as given the photon detection at $t$, the actually time $t'$ for the photon
staying inside the cavity is uncertain, and we have to sum up all the possible contributions from $0$ to $t$. This situation is illustrated in Fig.~\ref{fig:phasespace}.

One important feature in the above expression is that the integrand is a periodic function. If we denote
\be\label{eqfortime}
\tau \equiv \omega_m t = 2 \pi N+ \Delta \phi\,,
\ee
where $N$ is some integer and $\Delta \phi$ is the residual phase ranging from $0$ to $2\pi$. In this way, the integral in Eq.~(\ref{eqcft}) 
then becomes
\be\label{eqndp}
\left[N \int _0^{2\pi} +\int_0^{\Delta \phi}\right]{\rm d}\phi \,e^{-i n\phi}\ket{-\beta e^{i\phi}}
\ee
In the limit of $N\gg1$, when the photon arrives at the photodetector with a delay large compared to the oscillator's oscillation period, the first term in Eq.~\eqref{eqndp} always dominates.  This means we obtain the same conditional state if we restrict $\tau$ around an integer multiple of $2\pi$, or make sure it is large enough.  This leads to the interesting effect that in the asymptotic limit of $\tau \rightarrow +\infty$, the conditional state will be independent from $\tau$.  In practice, however, although the integral~\eqref{eqndp} increases with $N$, the exponential decay factor in Eq.~\eqref{eqcft} always favors simply choosing $N=1$.  It is straightforward to evaluate this conditional state; using 
\be
\int_0^{2\pi} {\rm d}\phi\, e^{-i n \phi} e^{e^{i\phi} \hat a^{\dag}}\ket 0\,= \frac{1}{n!}(\hat a^{\dag})^n\ket 0\,,
\ee
we have
\begin{eqnarray}
\label{integral}
\int _0^{2\pi}{\rm d}\phi\, e^{-i n\phi}\ket{-\beta e^{i\phi}}&=&\frac{2\pi (-\beta)^n e^{-\frac{\beta^2}{2}}}{\sqrt{n!}}\ket n  \nonumber\\&=&2\pi |n\rangle\langle n |-\beta\rangle\,, 
\end{eqnarray}
which means
\begin{equation}
|\psi\rangle_{\rm m} = \frac{ \pi \gamma^{3/2} e^{-\frac{\pi\gamma}{\omega_m}} e^{2\pi i \beta^2 }}{\omega_m } \frac{(-\beta)^n e^{-\frac{\beta^2}{2}}}{\sqrt{n!}} |\tilde n\rangle
\end{equation}
This is indeed proportional to $ |\tilde n\rangle$, as promised.  Here we have used
\begin{equation}
\langle -\beta |n\rangle =\frac{(-\beta)^n e^{-\beta^2/2}}{\sqrt{n!}}
\end{equation}

\begin{figure}
\includegraphics[width=0.475\textwidth]{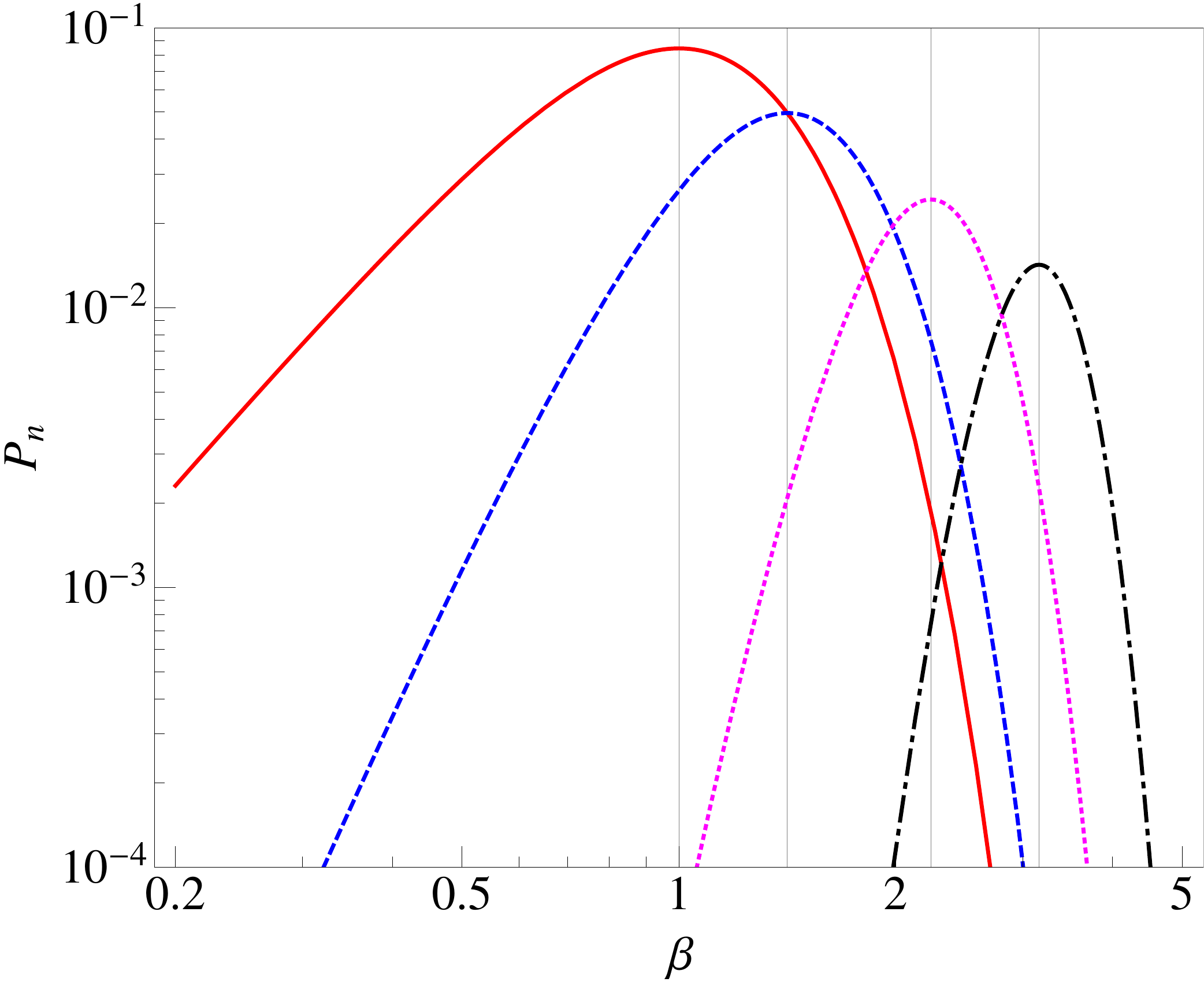}
\caption{Probability for obtaining displaced Fock states $|\tilde 1\rangle$ (red solid), $|\tilde 2\rangle $ (blue dashed), $|\tilde 5\rangle$ (magenta dotted) and $|\tilde 10\rangle$ (black dash-dotted), a range of $\beta$ and minimum state overlap of $1-\epsilon$. Vertical gridlines are draw for $\beta=1,\sqrt{2},\sqrt{5}$ and $\sqrt{10}$; these are the locations where maxima of $P_{1,2,5,10}$ are reached. \label{fig:singleprob}} 
\end{figure}

Since the probability for the returning photon to arrive at precisely $2\pi/\omega_m$ is zero, we must allow an interval around this target, which on the one hand provides us with a non-zero probability, but on the other hand makes the conditional state imprecise.  If we require the actual conditional state to have a high overlap with the target state (or high fidelity),
\begin{equation}
\frac{|{}_{\rm m}\langle \psi |\tilde n\rangle|}{\sqrt{ {}_{\rm m}\langle \psi |\psi\rangle_{\rm m}}}  \ge 1-\epsilon
\end{equation}
then, by perturbing the integration upper bound of Eq.~\eqref{integral}, we obtain the following requirement on the allowed photon arrival time
\begin{equation}
\label{errorarrival}
|\tau-2\pi|\le \Delta\tau \equiv \sqrt{8\pi^2\epsilon} \frac{|\langle -\beta | n\rangle|}{\sqrt{1-\langle -\beta | n\rangle^2}}\,,
\end{equation} 
which, for each trial of the experiment, would happen with a probability of
\begin{eqnarray}
\label{Pnsuccess}
P &=&\left|{}_{\rm m} \langle \psi |\psi\rangle_{\rm m}\right|^2 \frac{2\Delta \tau}{\omega_m}  \nonumber\\
&=& 2\sqrt{8\epsilon}\left(\frac{\pi\gamma}{\omega_m}\right)^3
 e^{-\frac{2\pi\gamma}{\omega_m}}\frac{|\langle -\beta | n\rangle|^3}{ \sqrt{1-|\langle -\beta | n\rangle|^2}} \,.
\end{eqnarray}
And this would be the probability with which we can create a conditional state with overlap at least $1-\epsilon$ with the target. 

From Eq.~\eqref{Pnsuccess}, we further notice that we should fix
\begin{equation}
\label{gammaomega}
\gamma/\omega_m =3/(2\pi)
\end{equation}
in order to obtain a maximized success probability of
\begin{equation}
\label{eqPbetan}
P_n =\sqrt{8\epsilon}\frac{27}{4e^3} \frac{|\langle -\beta | n\rangle|^3}{\sqrt{1-|\langle -\beta | n\rangle|^2}} 
\end{equation}
For each $n$, the maximum of $P_n$ is reached at $\beta=\sqrt{n}$. In Fig.~\ref{fig:singleprob}, we plot $P_n$ for a range of $\beta$, for $\epsilon=0.1$, or a state overlap of $\ge 90\%$.   We can see that the probability of producing $|\tilde n\rangle$ decreases rather quickly as $n$ increases.

This dependence \eqref{eqPbetan} on $\beta$ comes from two sources, which we can understand better by going to the phase-space reference frame  centered at the equilibrium position of the oscillator when the photon is inside the cavity.  In this reference frame, the complex amplitude of the coherent states being superimposed are located on a circle with distance $\beta$ away from the center, while the target we would like to prepare is simply the Fock state $|n\rangle$.  Although the photon's wave function selects out an oscillator state proportional to $|n\rangle$, this post-selection does not improve the intrinsic overlap between all those that participate the superposition, which is actually proportional to 
\begin{equation}
|\langle -\beta e^{i\phi} | n\rangle|^2  = |\langle -\beta  | n\rangle|^2  \,,
\end{equation} 
This explains the dependence of ${}_{\rm m}\langle \psi|\psi\rangle_{\rm m}$ on beta.  The other factor of dependence on $\beta$ is that when the target state has a very low overlap with the individual members $|\beta e^{i\phi}\rangle$ of the superposition, the requirement on the accuracy of photon arrival time, or $\Delta \tau$, increases, as shown in Eq.~\eqref{errorarrival}.

%Basically, in the long run, the phase driving term $e^{i n \omega_m t'}$ beats with the  freely evolving coherent state of the
%mechanical oscillator. The final state  asymptotically approaching a number state around a new equilibrium position, 
%and the deviation from this asymptotic state becomes very small after many oscillation cycles. Note that such an asymptotic 
%number state can also be achieved whenever $\Delta \phi=0$. 

\subsection{Preparation of an arbitrary state}

Since the displaced number states forms a complete basis we can expand any target state as
\be
\ket {\psi_{\rm tg}}=\sum _{n=0}^{+\infty} c_n \ket {\tilde n}\,,\quad \sum_{n=0}^{+\infty} |c_n|^2=1\,.
\ee
Since a linear combination of $F$'s leads to a linear combination of conditional states, we simply need to apply the result of the last subsection and have 
\be
\label{converg}
F(x)=  \frac{\sqrt{\gamma}e^{(\gamma/2-i\beta^2\omega_m)x}}{Z}  \displaystyle \sum_{n=0}^{+\infty}{ \tilde c_n e^{in\omega_mx}}
\ee
with
\begin{eqnarray}
Z&\equiv &{\left[\displaystyle  \sum_{j,k=0}^{+\infty} { \frac{ \tilde c_j \tilde c_k^* }{1+ i(j-k) \frac{\omega_m}{\gamma}}}\right]^{1/2}}\,,\\
\tilde c_n &\equiv & \frac{c_n}{\langle -\beta  |n \rangle} =\sqrt{n!} (-\beta)^n e^{\beta^2/2} c_n\,.
\end{eqnarray}
This is an additional periodic modulation (with period $2\pi/\omega_m$) of the photon's wave function. We caution that in order for the summation in Eq.~\eqref{converg} to converge, if $c_n$ does not go to zero for all $n \ge N$, then it must decay very fast when $n\rightarrow +\infty$, due to the presence of the $\sqrt{n!}$ factor (which grows faster than $\beta^{-n}$).

As in the previous subsection, we obtain the conditional state at $\tau \equiv \omega_m t = 2\pi, 4\pi, \ldots$, as well as any $\tau$ that is substantially large. Again, let us consider $\tau = 2\pi$, this gives the conditional state of
\begin{equation}
|\psi\rangle_{\rm m} = \frac{ \pi \gamma^{3/2} e^{-\frac{\pi\gamma}{\omega_m} } e^{2\pi i \beta^2 } }{\omega_m Z} \ket{\psi_{\rm tg}}
\end{equation}

\begin{figure*}
\includegraphics[width=0.7\textwidth]{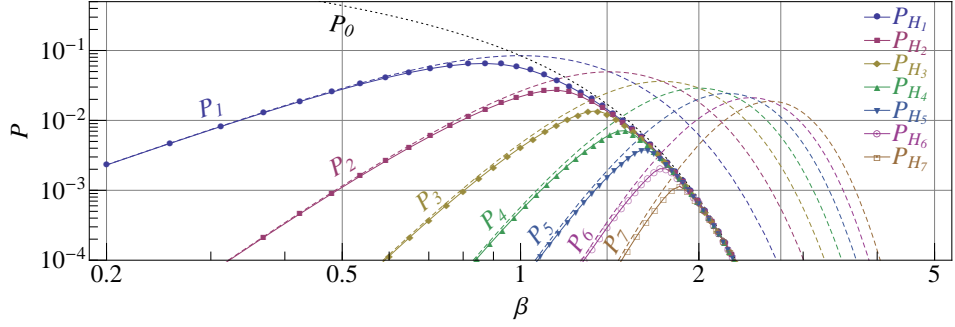}
\psfrag{H}{$\mathcal{H}$}
\caption{Minimum success probability for states in Hilbert spaces $\mathcal{H}_{1,2,\ldots 7}$ (solid curves with markers), together with success probability for producing single displaced Fock states, $P_{0,1,2,\ldots,7}$ (dashed curves without markers). Fidelity is fixed at $10\%$. Note that $P_0$ would become greater than 1 at low values of $\beta$ --- but in this case our approximation in obtaining $\Delta \tau$ breaks down. \label{fig:singleprob}  \label{fig:prob}} 
\end{figure*}

We can use the same approach as the previous subsection to evaluate the probability with which this conditional state is achieved with a high overlap.  For a minimum overlap of $1-\epsilon$, we require
\begin{equation}
|2\pi-\tau| \le \Delta \tau =\frac{\sqrt{8\pi\epsilon}}{ \left|\sum_{m=0}^{+\infty}\tilde c_m\right| \sqrt{1- |\langle -\beta|\psi_{\rm tg}\rangle|^2}}
\end{equation}
Note that this $\Delta\tau$ diverges if $\sum_{m=0}^{+\infty}\tilde c_m =0$, because in this case the overlap does not vary at $O[(\tau-2\pi)^2]$ order. Assuming the target state to be generic, then the probability for obtaining this state is then 
\begin{equation}
P_{|\psi\rangle} =  2\sqrt{8\epsilon} \frac{\left(\frac{\pi\gamma}{\omega_m}\right)^3 e^{-\frac{2\pi\gamma}{\omega_m}}\left[1- \left|\sum_{n=0}^{+\infty} \langle -\beta | n\rangle^2 \tilde c_n\right|^2\right]^{-1/2}}
{\left|\sum_{m=0}^{+\infty}\tilde c_m\right|\displaystyle \sum_{j,k=0}^{+\infty} \frac{\tilde c_j \tilde c_k^*}{1+i\frac{(j-k)\omega_m}{\gamma}}}
\end{equation}
Here the choice of $\gamma/\omega_m$ depends on the target quantum state, but if we assume this dependence is weaker than the pre-factor, and continue to use Eq.~\eqref{gammaomega}, then we obtain
\begin{equation}
P_{|\psi\rangle} =  \frac{27}{e^3} \sqrt{\frac{\epsilon}{2}} \frac{ \left[1- \left|\sum_{n=0}^{+\infty} \langle -\beta | n\rangle^2 \tilde c_n\right|^2\right]^{-1/2}}
{\left|\sum_{m=0}^{+\infty}\tilde c_m\right|\displaystyle \sum_{j,k=0}^{+\infty} \frac{\tilde c_j \tilde c_k^*}{1+\frac{2\pi i(j-k)}{3}}}
\end{equation}

As it turns out, $P_{|\psi\rangle}$ depends on the detail of $|\psi\rangle$ --- even if we only try to create a combination of $|\tilde 0\rangle$ and $|\tilde 1\rangle$, the combination coefficients would lead to very different success probabilities. In order to provide a concrete measure of the ability of our state-preparation scheme,  we have chosen to  compute the minimum success probabilities of creating all the states in the mechanical oscillator's Hilbert subspaces spanned by the lowest displaced Fock states, e.g.,  $\mathcal{H}_1 \equiv \mathrm{Sp}\{|\tilde 0\rangle, |\tilde  1\rangle\}$,  $\mathcal{H}_2 \equiv \mathrm{Sp}\{|\tilde 0\rangle, |\tilde 1\rangle, |\tilde 2\rangle\}$, etc.  We define
\begin{equation}
P_{\mathcal{H}_j} =\min_{|\psi\rangle \in \mathcal{H}_j} P_{|\psi\rangle}\,,\quad \mathcal{H}_j =
\left\{\sum_{l=0}^j \alpha_l |\tilde l\rangle:  \alpha_l \in \mathbb{C}\right\}
\end{equation}

In Fig.~\ref{fig:prob}, we plot $P_{\mathcal{H}_1}$, $P_{\mathcal{H}_2}$, \ldots, $P_{\mathcal{H}_7}$ as functions of $\beta$ (in solid purple curves).  Because $\mathcal{H}_1 \subset \mathcal{H}_2 \subset \cdots \subset \mathcal{H}_7$, it is increasingly difficult to create all states in $\mathcal{H}_j$ with higher values of $j$, and therefore $P_{\mathcal{H}_1} \ge P_{\mathcal{H}_2} \ge \ldots P_{\mathcal{H}_7}$, namely our success probability decreases globally when $j$ increases. In fact, as we overlay the single-Fock-state success probabilities $P_0$, $P_1$, \ldots, $P_5$, we also discover that for any $P_{\mathcal{H}_j}(\beta)$, it asymptotes to $P_0$ at higher $\beta$, and to $P_j$ at lower $\beta$; moreover, the transition between these two asymptotic regions are brief, and the $P_{\mathcal{H}_j}(\beta)$ curves do not lie much below the minimum of $P_0$ and $P_j$. 

This asymptotic behavior can be understood from the behavior of $P_n$, the success probability for single (displaced) Fock states.  For smaller $\beta$, it is much more difficult to prepare a higher Fock state, therefore, if $\beta$ is sufficiently small, the difficulty of preparing $\mathcal{H}_j$ is dominated by the preparation of  $|\tilde j\rangle$, the single most difficult state in the space to prepare --- and therefore $P_{\mathcal{H}_j}$ agrees with $P_j$.  Vice versa, for sufficiently large $\beta$, the difficulty of preparing $\mathcal{H}_j$ lies in the preparation of $\tilde 0\rangle$, and therefore $P_{\mathcal{H}_j}$ would agree with $P_0$.  The fast transition between the two extremes indicates that when trying to prepare states in $\mathcal{H}_j$, the difficulty either lies in $|\tilde 0\rangle$, or in $|\tilde j\rangle$, and only for a small region of $\beta$ the two difficulties might compete with each other --- while none of the intermediate states contribute to the difficulty of state preparation.  This is consistent with the relative locations of the $P_n$ curves in Fig.~\ref{fig:prob}: (i) for any $\beta$, $P_{1,2,\ldots, j-1}$ are always much greater than the minimum of $P_0$ and $P_j$, and (ii) as we move away from the $\beta$ at which $P_0$ and $P_j$ crosses each other, their discrepancy increases quickly.

 As a matter of practicality, we see that if we choose $\beta \approx 0.87$ the probability of achieving, with an overlap (or fidelity) above 90\%,  any superposition of $|\tilde 0\rangle$ and $|\tilde 1\rangle$  (i.e., any member of the subspace $\mathcal{H}_1$) is guaranteed to be above $6.3\%$.  On the other end, with a probability of at least $0.1\%$, we can produce all states in the 8-dimensional subspace of $\mathcal{H}_7$.

\section{Practical Considerations}

In order to realize such a state-preparation scheme,  we need to fulfill the following three requirements. 
The {\it first} requirement is that the series in Eq.~(\ref{converg}) be converging. This can be satisfied if
$\beta \geq 1$. To see what this means, we restore all the physical units:
\be
\beta = \frac{k/(2\omega_m)}{\sqrt {\hbar m\omega_m / 2}}=\left[\frac{\hbar\omega_0}{c}\right]\sqrt{\frac{2}{\hbar m\omega_m}}\left[2\omega_m\frac{L}{c}\right]^{-1}\,.
\ee
It characterizes the momentum kick of photon $\hbar \omega_0/c$ to the oscillator during one oscillation
period compared to the ground state momentum uncertainty $\sqrt {\hbar m\omega_m/2}$. The momentum kick
from the photon needs to be big enough to substantially change the mirror state.
The {\it second} requirement is that cavity bandwidth be smaller than the mechanical frequency 
\be
 \gamma<\omega_m.
\ee
This is because we need to wait at least several oscillation periods to approach the asymptotic state, and 
the photon should be long enough such that we have a finite probability for detecting photon at $t> \omega_m^{-1}$.

Combining the above two conditions, we obtain the following relation
\be
\label{nonlinear}
\frac{\lambda}{\mathcal{F}}<\sqrt{\frac{\hbar }{2 m \omega_m}}
\ee
where $\lambda$ is the optical wavelength of the photon, $\mathcal{F}$ is the cavity finesse. This means the cavity linear dynamical range much be less than the zero point uncertainty to realize the optomechanical nonlinearity. An alternative scheme has been proposed to make it more achievable experimentally ~\cite{Miao}.

The {\it third} requirement is that the thermal decoherence effect be small within one mechanical oscillation period, 
namely [cf. also Eq.\,(5) in Ref.~\cite{Marshall}]:
\be
Q > \frac{k T_{E}}{\hbar \omega_m},
\ee
where $Q$ is the mechanical quality factor of the oscillator and $T_{E}$ is the environmental temperature.
These three requirements can be achieved experimentally, e.g., the current setups shown in Refs.\,\cite{Teufel,Painter}
the one proposed in Ref.\,\cite{Thompson}.  

Finally, we require the the capability of generating a single photon with an arbitrary wave function with duration comparable to the mechanical oscillation frequency of the photon.  This is possible with cavity QED systems, as has been discussed by Ref.~\cite{Kimble1, Kimble2, Walther}. 

\section{Conclusions}\label{sec4}

We have presented an exact solution to the open quantum dynamics of an single-photon interferometer with a movable mirror. Since the photon number is preserved, we have been able to write the total wave function of photon as three components: incoming photon, inside-cavity photon and out-going photon. We analyzed the details of how 
the photon exchanges between the cavity mode and the external continuous field.

We studied the fringe visibility of the interferometer in a specific case by injecting a single photon with exponentially decaying profile and with the movable mirror initially prepared at the ground state.   This scheme has been proposed by Ref.~\cite{Marshall} to explore decoherence of a macroscopic oscillator, although in that proposal the photon has been assumed to start off from inside the cavity.  In the limit when the photon pulse is short (or $a \gg \gamma$), we did recover the result of Ref.~\cite{Marshall}, although our result deviates significantly when $a$ becomes comparable to $\gamma$.  We believe this is experimentally relevant, because in the case $a \gg \gamma$, the probability for the photon to exit from the detection port is very small, and therefore the experiment may suffer significantly from imperfections. 

%By exploiting the interference visibility of the outgoing photon at different time,
%we can investigate the quantum behavior of the movable mirror in the interferometer, also the corresponding probability of detection is calculated with
% different photon and cavity parameters, which indicates, to get complete destructive interference, we require $\beta$ at least to be 1, also the photon length (characterized by $a$) needs to be less than the cavity bandwidth.  To increase the probability of detection, the oscillator's frequency is better to be larger than the cavity bandwidth. 
%Our calculation shows that it would be challenging to observe the quantum decoherence with Penrose's setup since the detection probability would be quite low.
%The model and calculation we presented could be applied 
%to the realistic experiments in which the photon are usually injected from external single-photon sources, .

We have also studied the use of such  nonlinear optomechanical interactions to prepare the mechanical oscillator into an arbitrary quantum state --- similar to the proposal of Ref.~\cite{Bose}, although not having to require that the photon to start off from within the cavity.  To realize this, we require that: (i)  the optomechanical cavity must be working in the nonlinearity regime [i.e., the cavity's spatial line width must be less than the oscillator's zero-point position fluctuation, see discussions above Eq.~\eqref{nonlinear}], (ii) the cavity's frequency width must be less than the mechanical oscillator's angular frequency, (iii)  the thermal decoherence time must be less than several times the mirror's period of oscillation, and (iv) we must be able to engineer the single-photon wave function arbitrarily, at a time scale comparable to the mirror's oscillation period and with coherence time longer than the cavity storage time.  Although we have shown mathematically that all quantum states whose expansion coefficients in the displaced Fock states $|\tilde n\rangle$ drop sufficiently fast as $n\rightarrow +\infty$ can be prepared by modulating the wave function of the incoming photon and conditioning over the arrival time of the returning photon,  in practice we will be confined to the superposition of a handful of nearby displaced Fock states.

\acknowledgements
We thank our other colleagues in the LIGO MQM discussion group for fruitful discussions. 
We acknowledge funding provided by the Institute for Quantum
Information and Matter, an NSF Physics Frontiers Center with support of
the Gordon and Betty Moore Foundation.  This work has also been supported by NSF grants PHY-0555406, PHY-0956189, PHY-1068881, as well as the David and Barbara Groce startup fund at Caltech.

\end{document}